\documentclass{article} 
\usepackage{iclr2025_conference,times}

\usepackage{subcaption}
\usepackage{wrapfig}
\usepackage{lipsum} 
\usepackage{makecell}
\usepackage{wrapfig}
\usepackage[fontset=fandol]{ctex}
\usepackage[english]{babel}

\usepackage{amsmath,amsfonts,bm}
\usepackage{multicol}

\def\eqref#1{equation~\ref{#1}}

\def\1{\bm{1}}

\DeclareMathAlphabet{\mathsfit}{\encodingdefault}{\sfdefault}{m}{sl}
\SetMathAlphabet{\mathsfit}{bold}{\encodingdefault}{\sfdefault}{bx}{n}

\usepackage{booktabs}      
\usepackage{multirow}      
\usepackage{amssymb}       
\usepackage{graphicx}
\usepackage{tablefootnote}
\usepackage{threeparttable}
\usepackage[export]{adjustbox} 
\usepackage{color}

\usepackage{newtxtext,newtxmath}
\usepackage{setspace}
\usepackage{colortbl}
\usepackage{xcolor}

\usepackage{hyperref}
\usepackage{url}
\usepackage{multicol}
\usepackage{aliascnt}
\usepackage{adjustbox}
\usepackage{siunitx} 
\usepackage{booktabs}
\usepackage{multirow} 
\usepackage{enumitem}
\usepackage{booktabs} 
\usepackage{amssymb}  
\usepackage{amsmath}  
\usepackage{graphicx} 
\usepackage{longtable} 
\usepackage{graphicx}
\usepackage{subcaption}
\usepackage{calc}
\usepackage[most]{tcolorbox}
\usepackage{listings}
\usepackage{tabularx}
\usepackage{multirow}
\usepackage{array}
\usepackage{booktabs}

\usepackage[most]{tcolorbox}
\usepackage{xcolor}

\definecolor{uclablue}{rgb}{0.15, 0.45, 0.68}
\hypersetup{
    breaklinks,
    citecolor=uclablue,
    colorlinks=true,
    urlcolor=blue
}

\newtcolorbox{AIbox}[2][]{aibox,title=#2,#1}
\tcbset{
  aibox/.style={
    top=10pt,
    colframe=black,
    colbacktitle=black,
    coltitle=white,
    center,
  }
}

\lstdefinelanguage{prompt}{
    basicstyle=\scriptsize\ttfamily, 
    mathescape=true,        
    escapebegin=\color{latentcolor},  
    escapeend={},
    escapechar=@,
    stringstyle = \color{myorange},
    showstringspaces = false,
    moredelim = [s][\color{mypink}]{`}{`},
    moredelim = [s][\color{mybrown}]{```json}{```},
    moredelim = [s][\color{latentcolor}]{<StartOfLatent>}{<EndOfLatent>},
    literate = %
        {\ \ a.\ }{{\textcolor{mypurple}{\ \ a.\ }}}5
        {\ \ b.\ }{{\textcolor{mypurple}{\ \ b.\ }}}5
        {\ \ c.\ }{{\textcolor{mypurple}{\ \ c.\ }}}5
        {\ \ d.\ }{{\textcolor{mypurple}{\ \ d.\ }}}5
        {\ \ e.\ }{{\textcolor{mypurple}{\ \ e.\ }}}5
        {\ \ f.\ }{{\textcolor{mypurple}{\ \ f.\ }}}5
        {\ \ g.\ }{{\textcolor{mypurple}{\ \ g.\ }}}5
        {\ \ h.\ }{{\textcolor{mypurple}{\ \ h.\ }}}5
        {\ I.\ }{{\textcolor{mypurple}{\ I.\ }}}4
        {\ II.\ }{{\textcolor{mypurple}{\ II.\ }}}5
        {\ III.\ }{{\textcolor{mypurple}{\ III.\ }}}6
        {\ IV.\ }{{\textcolor{mypurple}{\ IV.\ }}}5
        {\ V.\ }{{\textcolor{mypurple}{\ V.\ }}}4
}

\newtcblisting[auto counter, number within=subsection]{prompt}[4][]{%
  before skip=6pt, after skip=6pt,    
  top=6pt, bottom=6pt, left=7pt, right=7pt, 
  enhanced, breakable,
  colback=white, colframe=gray!65,
  boxrule=0.6pt, arc=2pt,
  drop shadow={black!15},             
  listing only,
  listing options={
    language=prompt,
    upquote=true,
    basicstyle=\scriptsize\ttfamily \setlength{\baselineskip}{1.1\baselineskip},
    columns=fullflexible,
    breaklines=true,
    breakindent=0pt,
    xleftmargin=\ifx\empty#2\empty0pt\else#2\fi,
    xrightmargin=\ifx\empty#3\empty0pt\else#3\fi,
    aboveskip=0pt,                    
    belowskip=0pt,                    
    showstringspaces=false,
  },
  title={\ifstrempty{#4}{Prompt \thetcbcounter}{Prompt \thetcbcounter: #4}},
  colbacktitle=gray!6, coltitle=black,
  attach boxed title to top left={yshift=-1mm, xshift=6pt},
  boxed title style={colback=gray!6, boxrule=0pt, sharp corners},
  #1
}

\newtcblisting[auto counter, number within=subsection]{showcase}[4][]{%
  before=\par\vspace{\baselineskip},
  after=\par,
  width=\linewidth,
  enhanced,
  arc=0em,
  boxrule=1pt,
  listing only,
  listing options={
    language=prompt,
    upquote=true,
    basicstyle=\scriptsize\ttfamily \setlength{\baselineskip}{1.1\baselineskip},
    breaklines=true,
    breakindent=0pt,
    xleftmargin=\ifx\empty#2\empty-12pt\else#2\fi,
    xrightmargin=\ifx\empty#3\empty-5pt\else#3\fi,
    aboveskip=-4pt,
    belowskip=-4pt,
    columns=fullflexible,
    },
  colback=white,
  colframe=gray,
  colbacktitle=gray!5,
  coltitle=black,
  attach boxed title to top center={yshift=-3mm},
  box align=center,
  parbox=false,
  title={\ifstrempty{#4}{Example \thetcbcounter}{Example \thetcbcounter: #4}},
  #1
}

\newtcolorbox{titleabstractbox}{
    width=\textwidth, 
    colback=cyan!30!white!10, 
    colframe=white, 
    boxrule=0pt, 
    arc=20pt,   
    left=-25pt, 
    right=-25pt, 
    top=15pt, 
    bottom=10pt, 
    before skip=15pt, 
    after skip=20pt 
}

\definecolor{linkColor}{rgb}{0.2,0.4,0.6}
\definecolor{myblue}{HTML}{0379AC}
\definecolor{myred}{HTML}{A50E50}
\definecolor{myorange}{RGB}{238, 133, 74}
\definecolor{latentcolor}{named}{cyan}
\definecolor{normalcolor}{RGB}{0, 0, 0}
\usepackage{marvosym}

\title{Covo-Audio Technical Report}

\author{
\Large
\textbf{Tencent} \\
\normalsize
}

\begin{document}
\begin{titleabstractbox}
\maketitle
\let\oldthefootnote\thefootnote

\let\thefootnote\oldthefootnote

\newtcolorbox{abstractbox}{
    width=\textwidth,
    colback=blue!5!white,       
    colframe=blue!50!black,     
    arc=3pt,                    
    boxrule=0.5pt,              
    left=15pt,                  
    right=15pt,                 
    top=10pt,                   
    bottom=10pt,                
    fonttitle=\bfseries,        
    coltitle=blue!50!black,     
    attach title to upper,      
    after title={\par\medskip}  
}

\begin{abstract}

End-to-end large audio language models (LALMs) have emerged as a promising paradigm for unifying speech processing and language intelligence, yet achieving strong semantic reasoning alongside high-quality audio interaction remains challenging. In this work, we present \textit{Covo-Audio}, a 7B-parameter end-to-end LALM that directly processes continuous audio inputs and generates audio outputs within a single unified architecture. Through large-scale curated pretraining and targeted post-training, \textit{Covo-Audio} achieves state-of-the-art or competitive performance among models of comparable scale across a broad spectrum of tasks, including speech-text modeling, spoken dialogue, speech understanding, audio understanding, and full-duplex voice interaction. Extensive evaluations demonstrate that the pretrained foundation model exhibits strong speech–text comprehension and semantic reasoning capabilities on multiple benchmarks, outperforming representative open-source models of comparable scale. Furthermore, \textit{Covo-Audio-Chat}, the dialogue-oriented variant, demonstrates strong spoken conversational abilities, including understanding, contextual reasoning, instruction following, and generating contextually appropriate and empathetic responses, validating its applicability to real-world conversational assistant scenarios. \textit{Covo-Audio-Chat-FD}, the evolved full-duplex model, achieves substantially superior performance on both spoken dialogue capabilities and full-duplex interaction behaviors, demonstrating its competence in practical robustness. To mitigate the high cost of deploying end-to-end LALMs for natural conversational systems, we propose an intelligence–speaker decoupling strategy that separates dialogue intelligence from voice rendering, enabling flexible voice customization with minimal text-to-speech (TTS) data while preserving dialogue performance. Overall, our results highlight the strong potential of 7B-scale models to integrate sophisticated audio intelligence with high-level semantic reasoning, and suggest a scalable path toward more capable and versatile LALMs. We open-source \textit{Covo-Audio-Chat} model and the inference pipeline. 

\vspace{5pt}

\includegraphics[height=1.2em, valign=m]{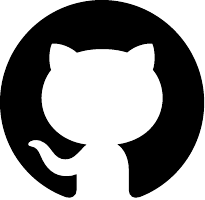} \hspace{0.3em}\textbf{GitHub:} \url{https://github.com/Tencent/Covo-Audio}

\includegraphics[height=1.2em, valign=m]{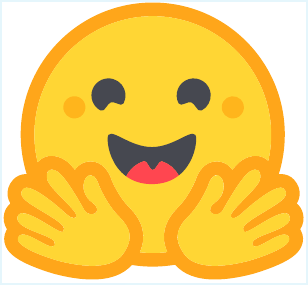} \hspace{0.2em}\textbf{HuggingFace:} \url{https://huggingface.co/tencent/Covo-Audio-Chat}



\end{abstract}
\end{titleabstractbox}

\section{Introduction}
\begin{figure*}[t]
	\centering 
	\includegraphics[width=\textwidth]{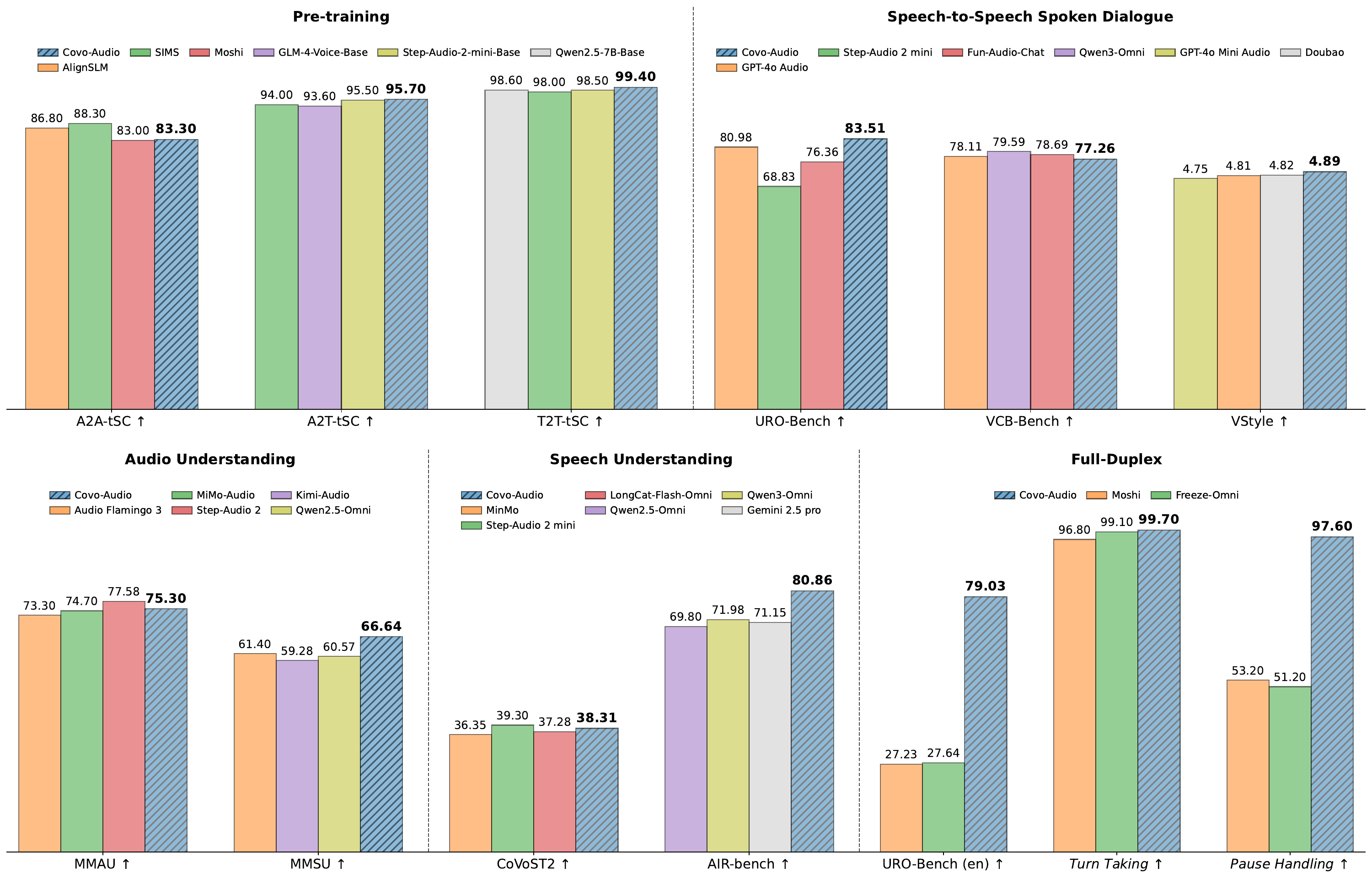} 
	\caption{An Overview of Comprehensive Performance Comparison.} 
    \label{results_overview}
\end{figure*}

Speech interaction, as the most natural and efficient form of human communication, is driving artificial intelligence toward more human-like conversational systems. In this context, the ideal conversational agent must harmonize intelligence (possessing deep linguistic understanding and reasoning), naturalness (preserving paralinguistic and emotional cues), and efficiency (enabling low-latency, fluent full-duplex voice interaction). However, current approaches often force compromises among these goals.

Traditional speech interaction systems predominantly rely on cascaded architectures, which combine independent modules for Automatic Speech Recognition (ASR), Large Language Model (LLM), and Text-to-Speech (TTS). While this modular paradigm offers interpretability and controllability, it suffers from the inherent issues such as information loss and error propagation, hindering real conversational experience. Recent large audio language models (LALMs) such as Qwen2.5-Omni \citep{xu2025qwen2} and Qwen3-Omni \citep{xu2025qwen3} adopt the Thinker-Talker architecture, where an intermediate textual reasoning step (the Thinker) precedes acoustic token prediction (the Talker). While this improves textual intelligence preservation, it sacrifices end-to-end speech instruction following abilities and direct conversational controllability. Additionally, handling full-duplex dynamics in such a sequential generation paradigm is more challenging.

Human voice interaction derives its richness from a dynamic blend of flexibility and diversity. End-to-end LALM, aiming to map raw audio input directly to audio output within a single, unified model, represents a promising frontier.
Pioneering works like GPT-4o \citep{hurst2024gpt}, GLM-4-Voice \citep{zeng2024glm} and Step-Audio 2 \citep{wu2025step2} have demonstrated the feasibility of this approach by augmenting LLMs with discrete audio tokens, showcasing the profound potential that enables low-latency, contextually consistent, and emotionally empathetic interaction. Architecturally in terms of decoding text-speech tokens, end-to-end models can be categorized into generating either interleaved streams \citep{wu2025step2, coreteam2025mimoaudio} or parallel streams \citep{ding2025kimi, chen2025fun, meituanlongcatteam2025longcatflashomnitechnicalreport}. We adopt the former as our backbone text-speech decoding design since it is more adherent to the oracle LLM philosophy. However, a common pain point accompanied with this unified approach is the deep \textbf{intelligence-speaker coupling problem} when developing a production-level application, which brings challenges in data preparation and deteriorates flexible voice customization. Building a chat bot combining virtues of fascinating voice and high-intelligence is rather laborious, since it is required to gather much high-quality dialogue data for the desired speaker\footnote{Commonly, we utilize advanced TTS engines to synthesize dialogue data, which inevitably lowers the upper bound of voice expressiveness.}. In this work, to mitigate this issue, we propose a novel \textbf{intelligence-speaker decoupling technique}, allowing us to build conversational capabilities directly from genuine TTS data, thereby avoiding the process of constructing lots of elaborate dialogue data. 


Another emerging research direction is end-to-end full-duplex LALMs, which facilitate human-like interaction with low latency and complex behaviors inherent to natural conversations. Representative works are Moshi \citep{defossez2024moshi} and Voila \citep{shi2025voilavoicelanguagefoundationmodels}, which typically employ a synchronous dual-stream scheme, where both input and output streams are in discrete token sequences, to achieve full-duplex. However, they still require word-level text-speech alignment to generate coherent speech. OmniFlatten \citep{zhang2025omniflattenendtoendgptmodel} adapts a text LLM backbone into a robust dialogue model through a curated multi-stage post-training. In this work, we put full-duplex interaction, without requirements of fine-grained text-speech alignment, directly into the pre-training phase to acquire robust full-duplex conversational capabilities via large-scale pre-training. Moreover, differing from other full-duplex models, we adopt a hybrid dual-stream scheme (i.e., continuous input stream and discrete output stream) in accordance with our half-duplex paradigm, which provides a more efficient and lossless way to perceive user expression and intention.



To facilitate these capabilities in an end-to-end paradigm, we present Covo-Audio, a compact LALM that achieves the fundamental comprehensive advantages of intelligence, naturalness, and efficiency that a voice conversational assistant should be empowered. Rather than focusing on a single task or setting, we demonstrate how pre-training and post-training strategies can endow an end-to-end LALM with robust audio perception, reasoning, emotional empathy, while economical flexible voice interaction capabilities. We conduct comprehensive evaluations of Covo-Audio across a broad range of tasks, including speech–text cross-modal alignments, speech understanding, audio question answering, and both half-duplex (Covo-Audio-Chat) and full-duplex (Covo-Audio-Chat-FD) speech-to-speech spoken dialogue. The results show that Covo-Audio achieves state-of-the-art (SOTA) or competitive performance among models of comparable scale. Our primary contributions are summarized as follows:


\textbf{Hierarchical Tri-modal Speech-Text Interleaving}: A framework designed to achieve deep alignment and fusion across modalities and scales. The Tri-modal aspect integrates continuous acoustic features, discrete speech tokens, and natural language text within a unified sequence, effectively bridging the gap between high-fidelity prosodic nuances and robust semantic structures. Hierarchical strategy aims to ensure structural coherence: phrase-level interleaving facilitates fine-grained alignment between acoustic fragments and lexical units, while sentence-level interleaving preserves the global semantic integrity and prosodic flow of long-form utterances. This dual-scale approach ensures the model captures precise acoustic nuances while maintaining macro-level coherence, allowing it to conform naturally to the LLM's generative patterns.


\textbf{Mitigating Intelligence-Speaker Coupling}: We propose a intelligence-speaker decoupling technique that decouples speaker from dialogue intelligence via multi-speaker training, then develop a contextual adaptation method to transfer and share high-quality TTS voices. By reformatting TTS recordings into pseudo-conversations with masked text loss, we preserve reasoning abilities while achieving high-fidelity voice naturalness, enabling versatile and personalized interaction in a more economical way.

\textbf{Native Full-Duplex Voice Interaction}: We evolve Covo-Audio-Chat into Covo-Audio-Chat-FD, a variant with low-latency, fluent full-duplex capability. In a more native way, we put full-duplex interaction directly into the pre-training phase to acquire full-duplex conversation capabilities via large-scale pre-training. Through large-scale pre-training and elaborate lightweight supervised fine-tuning, it supports robust real-time conversational dynamics, including smooth turn-taking, pause handling, user interruption (barge-in), and backchanneling, while maintaining competitive performance with the half-duplex model.

\textbf{Comprehensive State-of-the-Art Performance}: Covo-Audio delivers consistently strong and competitive results across core speech and audio tasks, including speech understanding, audio question answering, and both half-duplex and full-duplex speech-to-speech dialogue, despite its compact 7B-parameter footprint. It substantially outperforms existing models of comparable size and, notably, even matches the performance of significantly larger systems on multiple key metrics. This demonstrates that careful co-design of architecture, multi-stage pre-training, and alignment-focused post-training enables high-fidelity and robust speech interaction without parameter bloat, providing a practical solution for efficient deployment.



\section{Methodology}

\subsection{Architecture}

\begin{figure*}[t]
	\centering 
	\includegraphics[scale=0.55]{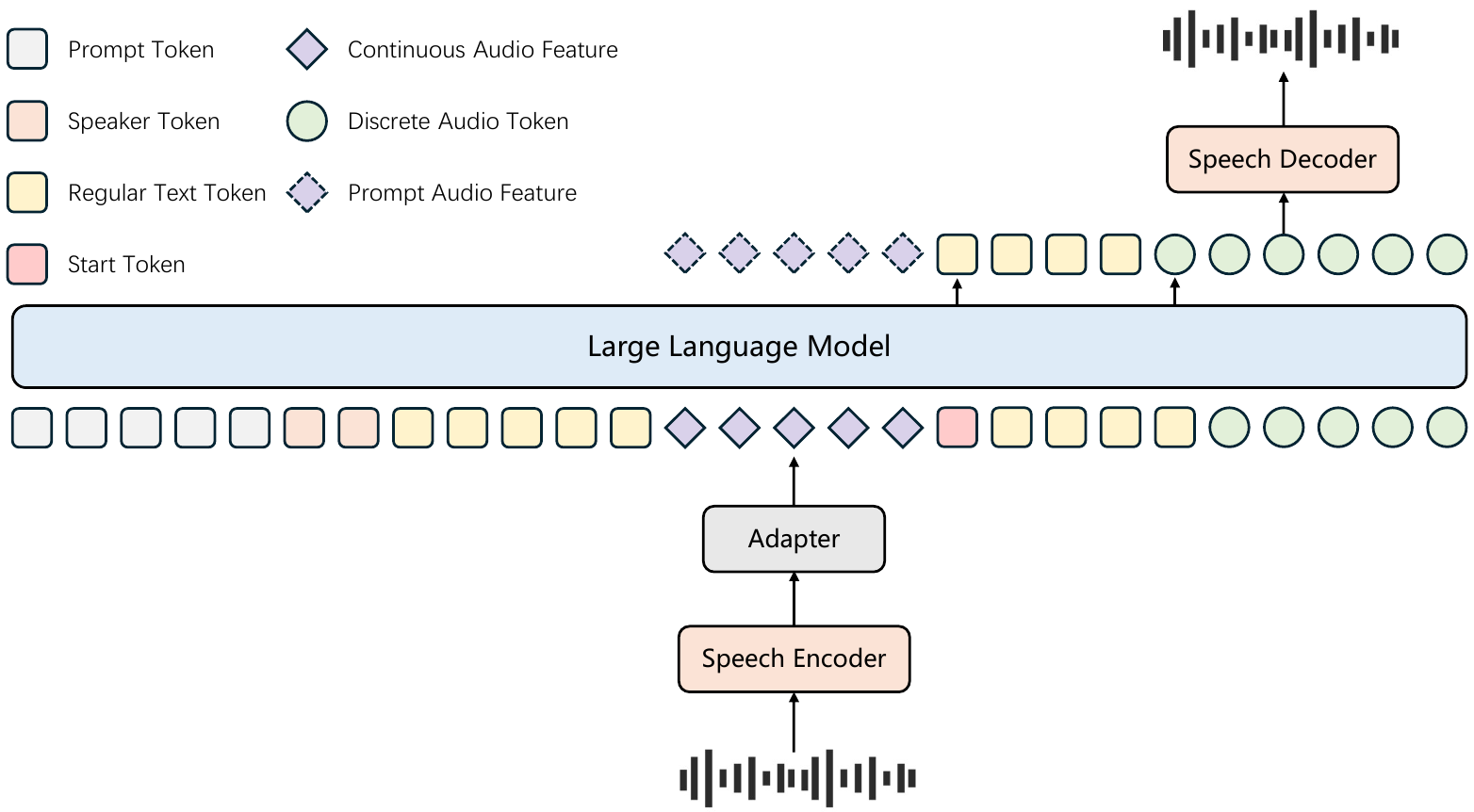} 
	\caption{An Overview of Covo-Audio.} 
    \label{framework_overview}
\end{figure*}

Covo-Audio employs a well-designed end-to-end architecture designed for seamless cross-modal interaction, as illustrated in Figure \ref{framework_overview}. The model consists of four primary components: an audio encoder, a robust LLM backbone, a specialized speech tokenizer, and a high-fidelity speech decoder.

We utilize the pre-trained Whisper-large-v3 \citep{radford2023robust} as our primary audio encoder. Whisper stands out for its exceptional robustness to background noise and varied accents, having been trained on a massive, weakly supervised multilingual dataset, and also guaranteeing a versatile, general-purpose ability of performing speech understanding. The output frame rate of the audio encoder is 50 Hz. An audio adapter is employed to bridge the audio encoder and the LLM. This adapter consists of three downsampling modules, each of which integrates two linear layers and a convolution layer with a downsampling rate of 2. Through this architecture, the adapter effectively reduces the frame rate from 50 Hz to 6.25 Hz.


To facilitate audio generation via the LLM, we developed a specialized speech tokenizer based on the WavLM-large \citep{chen2022wavlm}, augmented with a single Vector Quantization (VQ) layer \citep{van2017neural}. The tokenizer utilizes a codebook size of 16,384, producing discrete audio tokens at a rate of 25 Hz to ensure a high-compression yet information-rich representation. To achieve acoustic-semantic alignment, the tokenizer is pre-trained on a massive multilingual dataset. The training objective is guided by a multi-task loss function, which incorporates ASR loss for semantic grounding, TTS reconstruction loss for acoustic fidelity, and a specialized pitch loss to preserve the essential prosodic and intonational nuances of human speech. 

The speech decoder employs a two-stage hierarchical decoding framework to reconstruct high-fidelity audio from discrete tokens \citep{zhang2025minimaxspeechintrinsiczeroshottexttospeech}. A Flow-Matching (FM) \citep{du2024cosyvoice2} based decoder first transforms the discrete speech tokens into continuous acoustic-rich latent representations. Subsequently, these latents are fed into a BigVGAN \citep{lee2022bigvgan} based vocoder to reconstruct 24K waveforms.


Covo-Audio is built upon the Qwen2.5-7B-Base \citep{qwen2.5}, which we adapt for end-to-end speech interaction by extending audio encoder for perception, and by expanding its original text vocabulary with discrete audio tokens for generation. The architecture is designed to natively process interleaved inputs, consisting of continuous acoustic features (processed via an audio encoder and adapter) and textual tokens, while generating a unified sequence of text and audio tokens as output. For multi-turn conversational scenarios, we implement a recursive context-filling strategy where the continuous audio features from the user's input, along with the model's generated text and audio tokens from previous turns, are prefixed to the current input as historical context.

\subsection{Pre-training} \label{Pre-training}

Covo-Audio is initialized from Qwen2.5-7B-Base and undergoes a rigorous two-stage pre-training pipeline, processing a cumulative total of 2T tokens. This process is specifically engineered to fuse speech and text modalities within the model's latent space, evolving it from a text-centric LLM into a unified multi-modal engine capable of seamless cross-modal understanding and generation

\begin{table*}[!t]
    \small
    \centering
    \caption{List of pre-training tasks. $x{\to}y$ denotes $x$ is query and $y$ response, $x|y$ denotes paired queries or responses. $a_c$, $a_d$, $t$ denote continuous audio, discrete audio, and text, respectively. The trained tokens are rounded for ease of presentation.}
    \label{pre-tasks}
    \resizebox{\linewidth}{!}
    {
    \begin{tabular}{ccllc}
    \toprule
    Training Stage & Training Module & Task & Formula & Trained Tokens\\
    \midrule
    Stage 1 & Adapter & ASR & $a_c{\to}t$ & 30B \\
    \midrule
    \multirow{6}{*}{Stage 2}   &\multirow{6}{*}{Adapter+LLM} & ASR & $a_c{\to}t$ & 80B \\
    &                              & TTS & $t{\to}a_d$ & 160B \\
    &                              & Audio-only& $a_c{\to}a_d$, $a_d{\to}a_d$ & 240B \\
    &                              & Speech Continuation& $a_c{\to}t$ & 160B \\
    &                              & \multirow{2}{*}{Speech-Text Interleave}& $a_c{\to}t|a_d$, $a_c{\to}t{\to}a_c$, & 360B\\
    & & & $a_d{\to}t{\to}a_d$, $a_c{\to}t{\to}a_d$ &180B \\
    &                              & Text-only& $t{\to}t$ & 800B \\
    &                              & Full-Duplex & $a_c{\to}t|a_d$ & 5B \\
    \bottomrule
    \end{tabular}
    }
    
\end{table*}

\textbf{Stage 1: Modality Bridging via ASR Alignment.} This stage aims to establish a functional bridge between the audio encoder and the LLM, achieving initial semantic alignment between speech and text modalities. In this phase, we keep both the audio encoder and the LLM backbone frozen, focusing exclusively on optimizing the audio adapter. This alignment is supervised using a massive 200,000-hour multilingual ASR dataset, which forces the adapter to map continuous acoustic features into the LLM's latent space effectively. The training is conducted over 50k steps, and employs a cosine learning rate scheduler with a peak learning rate of $1 \times 10^{-4}$ and 40k steps for warm-up, which ensures stable convergence. By the end of this stage, the model develops a robust foundational ability to ``understand" spoken language, providing a well-aligned multimodal representation for subsequent pre-training.

\textbf{Stage 2: Speech-Text Modality Fusion.} This stage is specifically designed to achieve superior cross-modal alignment between the speech and text domains, moving beyond basic feature mapping to deep semantic and acoustic synchronization. During this phase, both the audio adapter and the LLM backbone are jointly unfrozen and optimized to refine the model's unified representation space. We implement a diverse multi-task training objective that includes ASR, TTS, Audio-only modeling, Speech Continuation, and Speech-Text Interleave, alongside Text-only to preserve foundational linguistic knowledge. Additionally, to acquire robust full-duplex interaction capabilities, we also include Full-Duplex pre-training in this stage. The entire pre-training tasks are listed in Table \ref{pre-tasks}.

A cornerstone of our pre-training phase is the Hierarchical Tri-modal Speech-text Interleaving strategy. Our framework acknowledges the dual nature of speech modeling: continuous acoustic representations $a_c$ derived from the Whisper encoder and discrete acoustic tokens $a_d$ generated by the speech tokenizer. Consequently, our pre-training objective is not merely to align speech with text, but to achieve a three-way fusion between continuous features, discrete tokens, and linguistic text within a unified latent space. In detail, we designed several tri-modal sequence structures. Specifically, we employ two primary patterns: (1) Sequential Interleaving ($a_c{\to}t{\to}a_d$): Arranging continuous features ($a_c$), text ($t$), and discrete tokens ($a_d$) in a progressive chain; (2) Parallel Integration ($a_c{\to}t|a_d$): Aligning continuous features with a coupled text-discrete unit. These paradigms ensure the model learns to map high-fidelity acoustic nuances (continuous) to robust semantic units (discrete and text) simultaneously. For Hierarchical Interleaving, previous interleaving methods, such as those utilized in GLM-4-Voice, primarily operate at the word or character level. While effective for local modality alignment, such granular interleaving often sacrifices the semantic integrity of long-form utterances and is highly sensitive to tokenization or forced-alignment errors. To overcome these challenges, our approach introduces a multi-scale interleaving mechanism that incorporates utterance-level paradigms alongside phrase-level segments. By interleaving longer, semantically complete speech and text fragments, we preserve the global context and natural flow of language. 

The training is conducted using a sequence length of 8192 and a cosine learning rate scheduler with a peak learning rate of $3 \times 10^{-5}$, spanning a total of 500k steps and 20k steps for warm-up. 
Our pre-training data consists of approximately 8 million hours of diverse audio and speech, and a text corpus encompassing multiple domains, totaling 3T tokens.
This comprehensive approach ensures that the model achieves high-fidelity modal consistency, enabling seamless and accurate transitions between spoken and written language.

\subsection{Speech-to-Speech Spoken Dialogue}
\begin{table*}[!t]
    \small
    \centering
    \caption{List of post-training tasks.}
    {
    \begin{tabular}{lcl}
    \toprule
    Task & Ratio & Formula \\
    \midrule
    General Intelligence & 0.4 & $t{\to}t$, $t{\to}a_d$, $a_c{\to}t$, $a_c{\to}t|a_d$ \\
    
    Spoken Dialogue & 0.3 & $a_c{\to}t|a_d$ \\
    
    Speech Understanding & 0.1 & $a_c{\to}t$, $a_c|t{\to}t$, $a_c|a_c{\to}t$, $a_c|a_c{\to}t|a_d$ \\
    
    Speech Generation & 0.1 & $t{\to}a_d$ \\
    
    Audio Understanding & 0.1 & $t|a_c{\to}t$ \\
    \bottomrule
    \end{tabular}
    }
    \label{post-tasks}
\end{table*}

To incentivize strong intelligence, natural spoken expression, and empathetic interaction capabilities, we design a collaborative data construction and training strategy that integrates text instruction tuning, audio adaptation, and spoken dialogue alignment, ultimately obtaining the spoken chat model, \textit{Covo-Audio-Chat}. Given the fundamental importance of the core intelligence and logical reasoning capabilities of Covo-Audio-Chat, it is essential to incorporate high-quality instruction data used in text LLM domain. Specifically, to transfer intelligence to the audio modality, we design multiple cross-modal tasks including text-to-text (T2T), text-to-audio (T2A), audio-to-text (A2T) and audio-to-audio (A2A) tasks. Note that in the A2A task, the query typically consists of audio input only, while the answer includes both text response and the corresponding audio in an interleaved format. The involved post-training tasks are listed in Table \ref{post-tasks}.

\textbf{Intelligence}  We start with a 10M internal text instruction data for T2T training. This corpus is balanced across diverse domains, including commonsense reasoning, mathematics, coding, and specialized knowledge. A randomly sampled subset is then converted into parallel speech data via a high-fidelity neural TTS system, enabling T2A, A2T and A2A training. Speaker diversity is ensured by randomly sampling voice styles from thousands of distinct speech prompts, followed by WER-based filtering to maintain speech quality. This parallel text–audio construction paradigm facilitates effective knowledge transfer from T2T to audio-related tasks, and finally mitigates intelligence degradation when extending the model to the speech modality. Additionally, we apply Kullback-Leibler (KL) based on-policy model distillation during T2T training, where a stronger text-based LLM provides top-20 logits as soft targets to prevent intelligence degradation.

\textbf{Colloquialism} Beyond intelligence, spoken dialogue requires colloquialism and conciseness. To enhance colloquial expressiveness, we select a moderate-scale set of dialogue samples suitable for spoken interaction and rewrite assistant responses into a conversational style using a text-based LLM before TTS synthesis. Distinct system prompts are employed during training to explicitly differentiate colloquial and formal response styles, enabling flexible spoken expression without sacrificing general capabilities.

\textbf{Empathy} To model empathetic speech interaction, we further construct emotion-aware dialogue datasets spanning seven emotion categories: joy, anger, sadness, fear, disgust, depression, and surprise. Each dialogue pairs explicit user emotional cues with assistant responses exhibiting active listening and context-sensitive support. These dialogues are synthesized with emotionally aligned speech, allowing Covo-Audio-Chat to jointly learn semantic and acoustic nuances for empathetic response generation.

The training is conducted using a sequence length of 8192 and a cosine learning rate scheduler with a peak learning rate of $1 \times 10^{-5}$, spanning a total of 50k steps and 1,000 steps for warm-up. The involved tasks with respective training ratio are listed in Table \ref{post-tasks}. This recipe endows Covo-Audio-Chat with strong logical reasoning and natural spoken dialogue capabilities.

\subsection{Intelligence-Speaker Decoupling Technique}
\begin{figure*}[t]
	\centering 
	\includegraphics[scale=0.65]{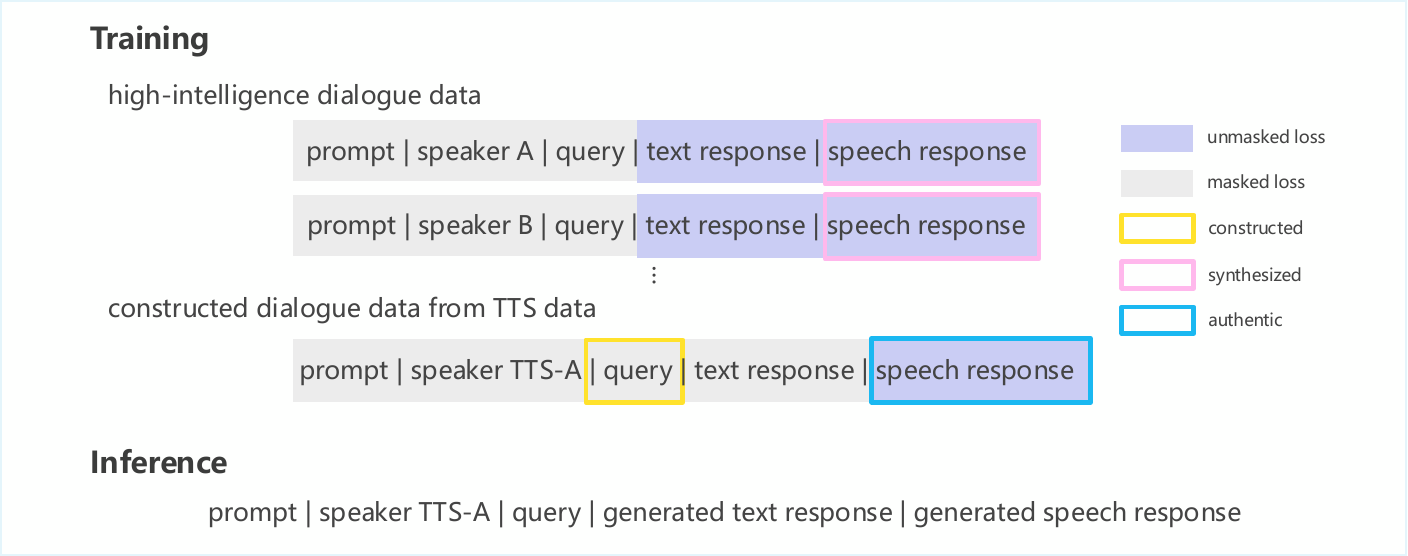} 
	\caption{Data Utilization in Intelligence-Speaker Decoupling Technique.} 
    \label{intelligence_speaker_decouping}
\end{figure*}

Building a versatile spoken dialogue model is rather challenging. There are two critical issues in end-to-end scenarios. (1) Due to the scarcity of authentic high-quality\footnote{That means of high intelligence and human-like.} spoken dialogue data, advanced TTS systems are often used to synthesize such data based on text dialogue datasets, not only inducing a prohibitive cost for flexible voice customization but also restricting the overall naturalness to the limitation of the TTS system. (2) On the other hand, high-quality TTS training data is relatively easy to collect but hard to exploit as it is originally not designed for dialogue-style interaction. A common practice is to construct conversational context manually based on such data, but often resulting in poor quality and degradation of overall dialogue performance. To address these two issues, during dialogue data preparation, we first randomly generated thousands of speakers to decouple speaker characteristics from intelligence, ensuring consistent intellectual levels across different speakers. However, when applying speakers trained solely for TTS tasks to dialogue tasks, we observed a significant decline in performance and reduced naturalness compared to TTS tasks. This drop is essentially due to the substantial format difference between TTS and dialogue tasks, making it difficult for models to directly transfer speakers across tasks. To further address this, we transformed high-quality TTS data into dialogue format via constructing appropriate pseudo-context, and incorporated it into training. To preserve dialogue intelligence effectively, we excluded the text response portion from the loss calculation. Experiments show that this approach successfully transfers the naturalness of TTS speakers while maintaining intellectual levels comparable to those of dialogue speaker. A data utilization schematic diagram is shown in Figure \ref{intelligence_speaker_decouping}. This technique enables versatile and personalized interaction while preserving high-fidelity voice naturalness with only lightweight TTS data needed.

\subsection{Full-Duplex Voice Interaction}
\begin{figure*}[t]
	\centering 
    \includegraphics[width=0.9\textwidth]{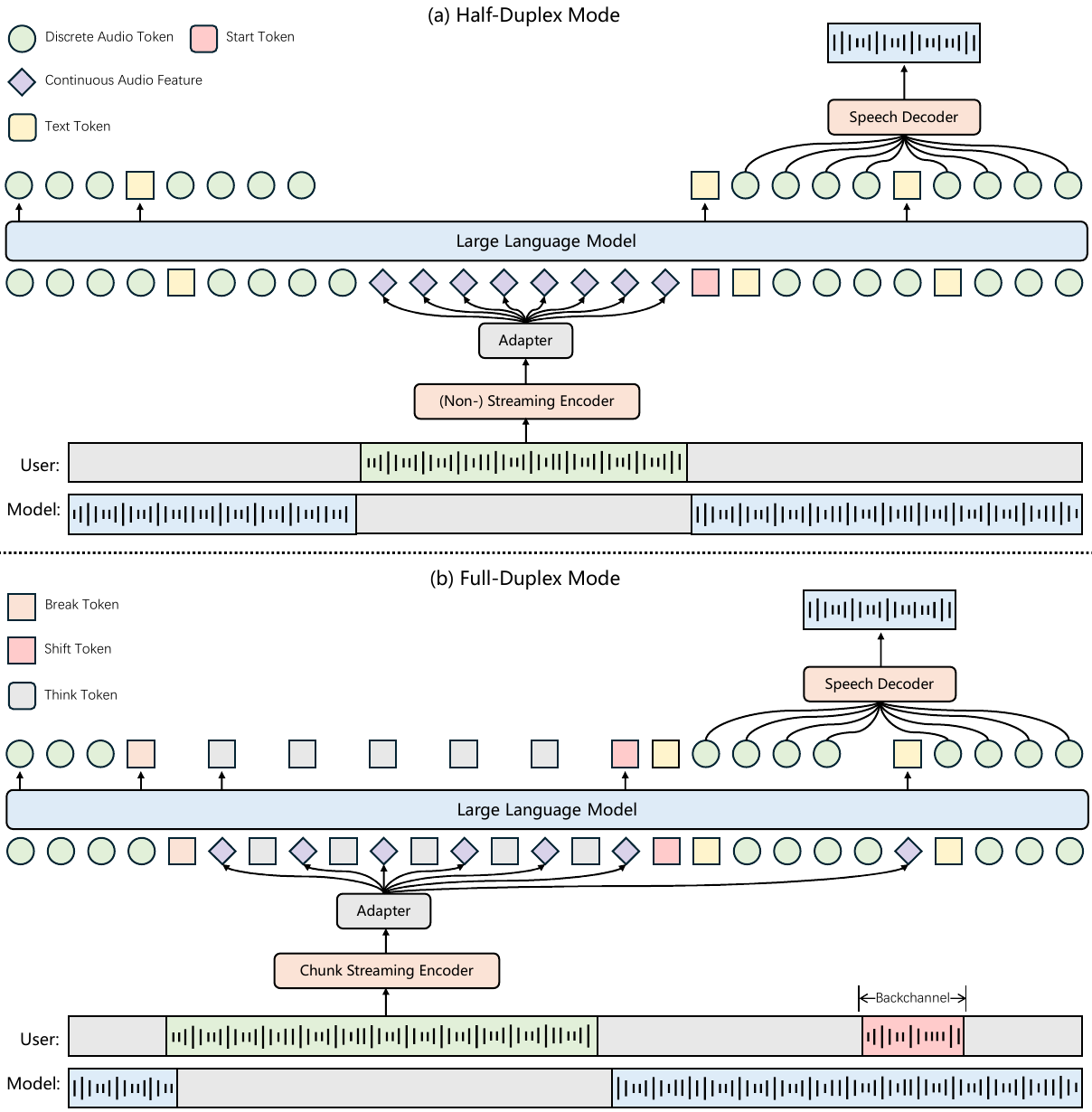} 
	\caption{A Comparative Overview of Covo-Audio-Chat and Covo-Audio-Chat-FD.} 
    \label{framework_hd_fd}
\end{figure*}

Covo-Audio-Chat still relies on turn-based conversation mechanism, where the model is in either listening or speaking state. In contrast, our full-duplex model, \textit{Covo-Audio-Chat-FD}, is endowed with simultaneous dual-stream communication that mirrors natural human interaction. To facilitate these merits, we made some modifications to the half-duplex model. First, the audio encoder is reformatted into a chunk streaming manner to enable real-time user stream input. Then the user stream and the model stream are chunk-interleaved in an 1:4 ratio\footnote{Since input audio features and output audio tokens are 6.25 Hz and 25 Hz respectively. Each chunk stands for 0.16s.} to allow the model to listen and speak simultaneously. A comprehensive comparison between Covo-Audio-Chat and Covo-Audio-Chat-FD is illustrated in Figure \ref{framework_hd_fd}. Additionally, we transformed the spoken dialogue datasets used in half-duplex training into dual-channel format and added conversational features such as barge-ins and backchanneling in proper positions according to semantic dynamics. To acquire robust full-duplex conversational capabilities, we performed large-scale pre-training, following the approach described in Section \ref{Pre-training}. Then we built Covo-Audio-Chat-FD directly from the pretrained foundation using both half-duplex and full-duplex datasets. The training is accomplished in one single step rather than using a multi-stage progressive training scheme as done in \citep{zhang2025omniflattenendtoendgptmodel}. We found this simple-yet-effective approach can yield more competitive results. In addition to intelligence preservation, full-duplex training also allows the model to master smooth turn-taking, to wisely handle user barge-ins and backchanneling behaviors.


Covo-Audio-Chat-FD works as follows. At start, during the user's turn, the model continuously outputs \texttt{THINK} tokens, indicating a listening-only state while it waits to respond appropriately. Upon the end of the user's turn, the model immediately outputs a \texttt{SHIFT} token to enter its speaking turn. Then four speech tokens are delivered sequentially preceded by a text token to guide semantic stability within each chunk, indicated as the listening-while-speaking state shown in Figure \ref{framework_hd_fd}(b). During this period, the model also receives the user stream simultaneously to facilitate flexible barge-ins or backchanneling. Once an interruption signal is detected, or otherwise it would end normally, the model immediately outputs a \texttt{BREAK} token and terminates speaking, switching to listening.

\subsection{Speech Understanding}
\label{speech_understanding}



Speech conveys diverse information, including both linguistic content and paralinguistic attributes such as speaker identity, emotional state, and demographic characteristics. To enable fluent speech interaction, we define the scope of speech understanding along two dimensions: (1) content understanding, which focuses on \textit{what is said}, and (2) attribute recognition, which captures \textit{how and by whom it is spoken}. Content understanding encompasses tasks including automatic speech recognition (ASR), speech-to-text translation (S2TT), and spoken language identification. Attribute recognition covers gender recognition, speaker counting, and emotion recognition. 

To equip the model with these capabilities, we construct multi-task question-answering data from large-scale speech corpora, including ASR datasets (LibriSpeech \citep{panayotov2015librispeech} etc.), speaker-rich datasets (VoxCeleb \citep{nagrani2017voxceleb}, CN-Celeb \citep{fan2020cn}), emotion-rich datasets (MELD \citep{meld}), multi-attribute datasets (Common Voice \citep{ardila2020common}), and speech reasoning subsets from AudioSkills \citep{AF3}. During data construction, we address two key challenges: (1) cross-modal alignment, which enables consistent speech understanding across input modalities, and (2) data bias mitigation, which ensures equitable model performance across demographic groups.

\textbf{Cross-Modal Data Construction}
We conduct cross-modal alignment to ensure the model can perceive paralinguistic attributes regardless of the input modality—whether the query is presented as text or speech.
Specifically, we design a unified data construction framework centered on speech recordings with specific attributes. Let $a^S$ denote the source speech containing target attributes, $t^Q$ and $a^Q$ the text query and speech query, and $t^R$ and $a^R$ the text response and speech response. We construct three complementary modality configurations:

\begin{itemize}
    \item \textbf{Text-to-Text QA}: $(a^S, t^Q) \rightarrow t^R$
    \item \textbf{Speech-to-Text QA}: $(a^S, a^Q) \rightarrow t^R$
    \item \textbf{Speech-to-Speech QA}: $(a^S, a^Q) \rightarrow (t^R, a^R)$
\end{itemize}

The construction pipeline proceeds as follows. First, we extract paralinguistic labels from open-source datasets and generate text-form QA pairs using predefined templates. Second, we synthesize corresponding speech queries and responses using high-quality TTS systems, ensuring natural prosody and diverse speaker characteristics. Finally, we mix all three modality configurations during training, enabling the model to learn modality-invariant representations for attribute recognition.

\textbf{Bias Mitigation}
Model bias has been a challenge in dialog systems, where models trained on imbalanced data often exhibit disparate performance across demographic groups, resulting in varying accuracy for different age, gender, or language groups. 
Such bias is particularly detrimental for production-level systems. 
To address this challenge, we employ stratified sampling to maximize demographic balance across paralinguistic attributes. Given the varying difficulty of balancing different attributes in available corpora, we adopt a hierarchical strategy with priority: age $\rightarrow$ language $\rightarrow$ gender, ensuring that the hardest-to-balance attributes receive primary consideration.

\subsection{Audio Understanding}

To cultivate robust audio understanding and complex reasoning capabilities, we employ a systematic training strategy centered on two distinct categories of data: Foundational Audio Perception Data and Cognitive Reasoning Data. These categories provide complementary signals that concurrently refine the model's acoustic grounding and analytical reasoning.

\textbf{Foundational Audio Skill Acquisition}
The first category focuses on establishing core acoustic-semantic grounding using the AudioSkill dataset \citep{goel2025audio}. This dataset encompasses a diverse array of tasks such as event recognition and cross-modal alignment, which are essential for mapping raw acoustic signals into the model's latent space. This process ensures that the model possesses the requisite perceptual ``vocabulary" to interpret complex auditory environments.

The training runs for 32k steps on the AudioSkill dataset containing 8M audio-text pairs. 
The model is optimized using AdamW~\citep{loshchilov2017decoupled} with differentiated learning rates: $3\times 10^{-6}$ for the LLM backbone and $1\times 10^{-5}$ for the audio adapter, with a weight decay of $5\times 10^{-4}$. We employ a linear warmup for $100$ steps followed by cosine annealing.

\textbf{Chain-of-Thought (CoT) Reasoning Enhancement}
The second category focuses on enhancing the model's reasoning capabilities for audio understanding tasks. Traditional audio language models often struggle with complex analytical tasks that require multi-step reasoning, such as emotion recognition involving subtle vocal cues, intent inference from prosodic patterns, or causal relationship identification in acoustic events. Chain-of-Thought (CoT) reasoning addresses this limitation by explicitly modeling the intermediate cognitive steps between audio perception and final conclusion, enabling the model to decompose complex audio comprehension into interpretable logical pathways.
To incorporate CoT capabilities, we construct a reasoning-enhanced dataset by merging and shuffling two complementary sources: AudioMCQ~\citep{he2025measuring} and AF-Think~\citep{goel2025audio}, totaling 1M training instances. All samples follow a unified format that explicitly separates the reasoning process from the final answer.

 
Since AF-Think originally provides only the reasoning process without explicit answers, we employ DeepSeek V3.1~\citep{deepseekai2024deepseekv3technicalreport} to extract and format the final answers, ensuring consistency across the entire dataset. We perform full-parameter training for two complete epochs using the same hyperparameters as the foundational stage, except with 1,000 warmup steps to accommodate the longer training duration. This CoT-enhanced training enables the model to not only recognize acoustic patterns but also articulate the reasoning behind its audio understanding decisions.

Following supervised fine-tuning, we employ Group Relative Policy Optimization (GRPO)~\citep{shao2024deepseekmath} to refine the model's audio reasoning capabilities using data from the Audio-Visual Question Answering (AVQA) benchmark~\citep{yang2022avqa}. Inspired by Audio-Thinker~\citep{wu2025audio}, which demonstrates the effectiveness of explicit reasoning for audio understanding, we design a verifiable composite reward function comprising four complementary components:
\begin{equation}
R_{\text{total}} = R_{\text{accuracy}} + R_{\text{format}} + R_{\text{consistency}} + R_{\text{thinking}}
\end{equation}
\noindent \textbf{Accuracy Reward} ($R_{\text{accuracy}} \in \{0, 1\}$): Measures whether the model's final answer matches the ground truth, providing direct supervision for correctness.

\noindent \textbf{Format Reward} ($R_{\text{format}} \in \{0, 1\}$): Verifies adherence to the structured output requirement, checking for the presence of both reasoning and answer sections:
\begin{equation}
R_{\text{format}} = \mathbb{1}\left[\text{match}(\text{output}, \texttt{<think>.*</think>.*<answer>.*</answer>})\right]
\end{equation}
\noindent \textbf{Consistency Reward} ($R_{\text{consistency}} \in \{0, 1\}$): Evaluates logical coherence between the reasoning process and final answer using an auxiliary LLM judge. This ensures that the model's explicit reasoning aligns with its conclusions.

\noindent \textbf{Thinking Quality Reward} ($R_{\text{thinking}} \in [0, 1]$): Assesses the reasoning depth across five dimensions: logical soundness, factual correctness, error identification, language coherence, and conciseness. The score is obtained via LLM-based evaluation in 0.2 increments.

This multi-faceted reward structure enables the model to simultaneously optimize for correctness, structural compliance, internal consistency, and reasoning quality, addressing limitations of supervised learning alone in capturing nuanced audio understanding behaviors.

\section{Experiments}
\subsection{Evaluation on Pre-training}

\begin{table*}[!t]
    \small
    \centering
    \caption{Performance comparison of pre-trained model and baselines on speech-text modeling capabilities.}
    \resizebox{\linewidth}{!}
    {
    \begin{tabular}{lccccccc}
    \toprule
    \multirow{2}{*}{Models} & \multirow{2}{*}{Params}& \multirow{2}{*}{A2A-tSC($\uparrow$)} &\multirow{2}{*}{A2T-tSC($\uparrow$)} &\multirow{2}{*}{T2T-tSC($\uparrow$)}  & \multirow{2}{*}{sBLIMP($\uparrow$)} & \multirow{2}{*}{sWUGGY($\uparrow$)} \\
    &&&&&&& \\
    \midrule
    Qwen2.5-7B-Base &7B &- &- &{98.6} &- &- \\
    \midrule
    TWIST &7B &74.1 &- &- &59.0 &73.90\\
    SpiritLM &7B &82.9 &88.6 &98.0 &58.3 &69.00\\
    Slam &358M &82.0 &- &- &58.8 &68.38 \\
    AlignSLM &7B &86.8 &- &- &\textbf{62.3} &-\\
    SIMS &7B &\textbf{88.3} &94.0 &98.0 &59.8 &\textbf{75.36} \\
    \midrule
    Moshi &7B &83.0 &- &- &58.8 &- \\
    MinMo &7B &75.4 &- &- &55.4 &68.59 \\
    LLaMA-Omni2 &7B &78.2 &- &- &53.6 &73.21 \\
    \midrule
    GLM-4-Voice-Base &9B &82.9 &93.6 &- &- &- \\
    Step-Audio-2-mini-Base &7B &- &95.5 &98.5 &52.4 &52.70 \\
    \midrule
    \textbf{\textit{Covo-Audio-Base}} &7B &83.3 &\textbf{95.7} &\textbf{99.4} &{61.6} &{74.90} \\
    \bottomrule
    \end{tabular}
    }
    \label{pt_exp_speech_text}
\end{table*}

\begin{table*}[!t]
    \small
    \centering
    \caption{Performance comparison of pre-trained model and baselines on ASR and TTS tasks. LS clean and LS other represent Librispeech test clean and test other set, respectively.}
    {
    \begin{tabular}{lccccccc}
    \toprule
    \multirow{2}{*}{Models} & \multirow{2}{*}{Params} &\multicolumn{3}{c}{ASR ($\downarrow$)}  &\multicolumn{2}{c}{Seed-TTS ($\downarrow$)} \\
    \cmidrule(lr){3-5} \cmidrule(lr){6-7}
    & &Aishell-1 &LS clean &LS other &Test-en &Test-zh \\
    \midrule
    Whisper-large-v3 &1.5B &9.31 &2.50 &4.53 &- &-\\
    CosyVoice2 &0.5B &- &- &- &2.57 &1.45 \\
    \midrule
    SpiritLM &7B &- &6.00 &11.0 &- &- \\
    
    GLM-4-Voice-Base &9B &2.46 &2.82 &7.66 &2.91 &2.10\\
    \midrule
    \textbf{\textit{Covo-Audio-Base}} &7B &1.96 &1.96 &4.55 &2.44 &1.73\\
    \bottomrule
    \end{tabular}
    }
    \label{pt_exp_asr_tts}
\end{table*}


To rigorously assess the multi-modal capabilities of our model, we established a comprehensive evaluation suite encompassing six distinct task categories. This framework covers A2A, A2T, and T2T scenarios. Specifically, for A2A tasks, we evaluated the model's ability to generate speech responses directly from speech input, focusing on dimensions such as creative story continuation (topic StoryCloze \citep{hassid2023textually}), grammatical integrity (sBLIMP \citep{nguyen2020zero}), and structural consistency (sWUGGY \citep{nguyen2020zero}). Based on the tSC data, we also evaluated the model's performance on both A2T and T2T tasks. Furthermore, standard benchmarks for ASR and TTS are integrated to provide a holistic measure of the model's foundational speech processing and synthesis performance.


To evaluate the performance of the pre-trained model, we compare it against a representative suite of speech language model baselines, categorized into three distinct groups. The first group consists of specialized A2A architectures, including TWIST \citep{hassid2023textually}, SpiritLM \citep{nguyen2025spirit}, Slam \citep{maimon2025slamming}, AlignSLM \citep{lin2025align}, and SIMS \citep{maimon2025scaling}. These models are specifically engineered and trained for direct A2A tasks. Secondly, we include Moshi, Minimo \cite{chen2025minmo} and Llama-Omni2 \cite{fang2025llama}, the prominent unified speech dialogue model recognized for its comprehensive modeling. Furthermore, we evaluated leading pre-trained foundations for speech-to-speech dialogue, specifically GLM-4-Voice-Base and Step-Audio2-mini-Base. Beyond cross-modal alignment performance, we extended our evaluation to include domain-specific models: Whisper is utilized for ASR benchmarks, while CosyVoice2 \cite{du2024cosyvoice2} served as the baseline for TTS. By benchmarking against this diverse array of models—ranging from specialized S2S, ASR, TTS frameworks to generalized pre-trained speech systems—we aim to demonstrate the capabilities in both high-level semantic reasoning and foundational acoustic generation.

Our analysis primarily focused on the efficacy of speech-text modality alignment. The results are shown in Table \ref{pt_exp_speech_text}. In the A2A story continuation task, our model achieved performance parity with sota pre-trained and specialized A2A models. It is noteworthy that many specialized A2A baselines, such as SpiritLM and Slam, exhibit signs of overfitting on continuation tasks, as they are often biased toward speech-to-speech mapping at the expense of broader linguistic flexibility. Furthermore, while models like AlignSLM and SIMS benefit from extra reinforcement learning, our results were achieved through pre-training alone. Crucially, our model significantly outperformed all existing baselines in logical reasoning, grammatical accuracy in A2A metrics, and cross-modality consistency across A2T and T2T benchmarks. \textbf{These results demonstrate that our pre-training objective successfully fused the speech modality with the core cognitive and generative powers of the LLM.} This suggests that the model does not merely transcribe audio but deeply extracts semantic information from speech signals to inform its high-level reasoning.

Beyond high-level reasoning, we evaluated the model's performance on fundamental ASR and TTS tasks. Despite being a unified multi-modal system, our model demonstrated competitive proficiency compared to the GLM-4-Voice-Base model, as shown in Table \ref{pt_exp_asr_tts}. The end-to-end pre-training paradigm we propose facilitates a robust bi-directional semantic mapping between speech and text. \textbf{This synergy ensures that the model preserves fine-grained acoustic details while maintaining the sophisticated linguistic understanding inherent in large-scale language modeling.}

\subsection{Evaluation on Speech-to-Speech Spoken Dialogue} \label{half_duplex_eval}
\textbf{Evaluation on Spoken Dialogue} We evaluate the end-to-end spoken dialogue capability of Covo-Audio-Chat on URO-Bench \citep{yan2025uro} and VCB Bench \citep{hu2025vcb} in an A2A mode. Specifically, we adopt the basic Chinese and English track of URO-Bench to assess the model's performance in speech comprehension, speech reasoning, and spoken dialogue. For VCB Bench, our evaluation focuses on the model's text-side and speech-side instruction-following ability, internal knowledge retention, and anti-interference capability. The models used for comparison include: GLM-4-Voice~\citep{zeng2024glm}, Freeze-Omni~\citep{wang2024freeze}, Kimi-Audio~\citep{ding2025kimi}, Qwen2.5-Omni~\citep{xu2025qwen2}, Baichuan-Audio-Chat \citep{li2025baichuan}, Step-Audio 2 series \citep{wu2025step2}, MiMo-Audio \citep{coreteam2025mimoaudio}, GPT-4o Audio, Qwen3-Omni \citep{xu2025qwen3}, and Fun-Audio-Chat \citep{chen2025fun}.

The results on the URO-Bench are shown in Table \ref{uro-all}. As illustrated, Covo-Audio-Chat demonstrates comprehensive superiority in core tasks, particularly excelling in speech reasoning and spoken dialogue which are critical for practical speech interaction. Specifically, on the Chinese track, Covo-Audio-Chat achieves the highest scores in four reasoning tasks (SQuAD: 77.34, OpenbookQA: 83.60, APE: 68.42, MLC: 80.69) and two spoken dialogue tasks (AlpacaEval: 90.02, Wildchat: 90.41), outperforming all comparative models including the advanced ones like Qwen3-Omni. In speech comprehension, its Repeat (98.17) and HSK5 (88.00) scores are among the top tiers, only slightly lower than Step-Audio 2 mini and MiMo-Audio respectively. On the English track, Covo-Audio-Chat maintains its strength in reasoning, achieving the best performance in Gsm8kEval (85.68). Though not leading in speech comprehension and partial spoken dialogue tasks, its overall performance is consistently competitive. In summary, \textbf{Covo-Audio-Chat demonstrates robust and balanced performance across both Chinese and English tasks despite at a compact 7B-parameter scale, highlighting its strong bilingual spoken dialogue capabilities}. This gives it a key advantage over many models that lack comprehensive bilingual coverage, and solidifies its overall dominance in core speech-to-speech interaction scenarios.

The results on the VCB Bench are shown in Table \ref{vcb}, evaluating speech-to-speech models across three core dimensions: Instruction Following, Knowledge, and Robustness. As illustrated, our proposed Covo-Audio achieves excellent performance in key metrics, demonstrating its superiority in practical speech interaction scenarios. In the Instruction Following task, Covo-Audio-Chat outperforms all competing models in Text-based Instruction Following (TIF: 93.07, TIF-En: 89.94) and Multi-Turn Dialogue (MTD: 87.70), highlighting its strong ability to comprehend and execute both Chinese and English text instructions, as well as maintain coherent and compliant communication in multi-round interaction. 
Notably, in the Robustness evaluation---critical for real-world applicability---Covo-Audio-Chat sets the state-of-the-art across all sub-tasks: Speaker Variation (SV: 88.94), Environmental Noise (EV: 87.13), and Content Variation (CV: 90.37), which underscores its robust adaptability to diverse interfering factors such as accents, background noise, and grammatical deviations. 
In the Knowledge dimension, Covo-Audio-Chat achieves competitive performance in Mathematical Logic (ML: 79.34), approaching the top-tier level, while its General Knowledge (GK) and Dialogue Comprehension (DC) scores are modest compared to Qwen3-Omni---this reflects a deliberate prioritization of core interaction capabilities during model development, and these aspects can be enhanced by integrating specialized knowledge bases and refining contextual modeling. Overall, \textbf{Covo-Audio-Chat demonstrates outstanding performance in most critical dimensions of speech-to-speech interaction (text instruction following, multi-turn dialogue, and robustness), laying a solid foundation for practical applications, with the identified limitations providing clear pathways for further optimization}.

\begin{table*}[!t]
    \captionsetup[subtable]{labelformat=simple}
    \small
    \centering
    \caption{\textbf{Performance Comparison on the Basic Track of URO Bench.} Results marked with ($^{*}$) are sourced from \citep{yan2025uro}, and those marked with ($^{\dagger}$) are from \citep{wu2025step2}. Covo-Audio-Chat-TTS denotes dialogue voice transferred from a TTS speaker.}
    \label{uro-all}

    \begin{subtable}{\linewidth}
        \centering
        \caption{Results on Chinese Basic Track.} 
        \label{uro-zh}
        
        \resizebox{\linewidth}{!}{
            \begin{tabular}{lcccccccccc}
            \toprule
            \multirow{2}{*}{Model}  &\multicolumn{3}{c}{Understanding($\uparrow$)} &\multicolumn{4}{c}{Reasoning($\uparrow$)} &\multicolumn{3}{c}{Oral Conversation($\uparrow$)} \\
            \cmidrule(lr){2-4}\cmidrule(lr){5-8}\cmidrule(lr){9-11}
            & Repeat & LCSTS & HSK5 & SQuAD & OpenbookQA & APE & MLC & AlpacaEval & Claude & Wildchat \\
            \midrule
            GLM-4-Voice$^{*}$ & 92.64 & 77.08 & 69.00 & 28.75 & 56.96 & 15.78 & 78.85 & 83.35 & 82.12 &84.48\\
            Freeze-Omni$^{*}$ & 4.97 & 71.82 & 7.66 & 9.58 & 16.40 & 11.75 & 47.35 & 67.98 & 64.89 & 71.28 \\
            GPT-4o Audio$^{*}$ & 93.50 & 81.60 & 88.00 & 42.67 & 76.00 & 25.33 & 81.33 & 86.40 & 82.93 & 80.00 \\
            Kimi-Audio$^{\dagger}$  & 73.32 & 78.40 & - & - &66.93 & - & 52.45 & 69.90 & 81.81 & - \\
            Qwen2.5-Omni$^{\dagger}$ & 19.66 & 73.22 & - & - & 67.70 & - & 61.76 & 70.40 & 79.71 & - \\
            Step-Audio 2$^{\dagger}$  & 96.16 & 79.16 & - & - & 74.06 & - & 62.99 &76.23 & 84.54 & - \\
            Step-Audio-AQAA$^{\dagger}$  &63.72 &68.33& - & - &47.21 & - & 67.40 & 56.23 & 31.50 & - \\
            \midrule
            Step-Audio 2 mini  & \textbf{99.83} & 81.18 & 80.67 & 64.71 & 70.02 & 45.09 &74.94& 74.29& 72.58&79.58 \\
            MiMo-Audio  & 0.44 & 79.61 & \textbf{92.67} & 35.73 &\textbf{83.60} & 61.05 & 38.16 & 85.90 & \textbf{89.10} & 84.84 \\
            Qwen3-Omni  & 99.48 & \textbf{84.54} & 88.67 & 52.07 & 73.02 & 26.67 & 36.09 & 79.59 & 78.86 & 74.54 \\
            Fun-Audio-Chat  & 97.50 & 80.95 & 79.00 & 51.85 & 67.90 & 44.91 & 76.09 & 80.59 & 86.46 & 84.62 \\
            \midrule
            \textbf{\textit{Covo-Audio-Chat}} & 98.17& 79.89& 88.00& \textbf{77.34}& \textbf{83.60}&\textbf{68.42} & \textbf{80.69}&\textbf{90.02} & 87.18& \textbf{90.41}\\
            
            \textbf{\textit{Covo-Audio-Chat-TTS}} & 97.68& 79.22& 91.00& 76.91& 80.60& 64.21& 79.31& 89.34& 87.99& 89.28\\
            \bottomrule
            \end{tabular}
            }
    \end{subtable}

    \vspace{1em} 

    \begin{subtable}{\linewidth}
        \centering
        \caption{Results on English Basic Track.} 
        \label{uro-en}
        
        \resizebox{\linewidth}{!}{
            \begin{tabular}{lcccccccccc}
            \toprule
            \multirow{2}{*}{Model}  &\multicolumn{3}{c}{Understanding($\uparrow$)} &\multicolumn{4}{c}{Reasoning($\uparrow$)} &\multicolumn{3}{c}{Oral Conversation($\uparrow$)} \\
            \cmidrule(lr){2-4}\cmidrule(lr){5-8}\cmidrule(lr){9-11}
            & \scriptsize{Repeat} & \scriptsize{Summary} & \scriptsize{GaokaoEval} & \scriptsize{StoralEval} & \scriptsize{TruthfulEval} & \scriptsize{Gsm8kEval} & \scriptsize{MLC} & \scriptsize{AlpacaEval} & \scriptsize{CommonEval} & \scriptsize{WildchatEval} \\
            \hline
            GLM-4-Voice$^{*}$ & 90.95 & 91.07 & 64.47 & 73.80 & 59.28 & 30.93 & 57.82 & 80.77 & 63.07 & 78.76\\
            Freeze-Omni$^{*}$ & 70.89 & 78.87 & 26.29 & 57.74 & 46.95 & 2.81 & 42.56 & 52.23 & 48.70 & 55.80 \\
            GPT-4o Audio$^{*}$ & 97.16 & 94.13 & 72.00 & 84.27 & \textbf{82.67} & 80.00 & 80.00 & \textbf{95.20} & \textbf{94.13} & \textbf{95.20} \\
            Kimi-Audio$^{\dagger}$  & 90.27 & 78.19 & 81.63 & 70.88 & 53.90 & 0.97 & 43.50 & 63.05 & 51.40 & 66.63 \\
            Qwen2.5-Omni$^{\dagger}$ & 28.36 & 80.51 & 89.99 & 72.90 & 62.03 & 63.52 & \textbf{80.04} & 78.16 & 72.40 & 77.92 \\
            Step-Audio 2$^{\dagger}$  &97.74&83.90&\textbf{90.76}&76.38&65.45&62.83&77.02&81.74&72.87&81.62\\
            Step-Audio-AQAA$^{\dagger}$  & 93.02 & 90.62 & 86.80 & 53.00 & 62.30 & 38.20 & 71.00 &71.86 & 75.53 &68.79  \\
            \midrule
            Step-Audio 2 mini & \textbf{97.81} & 84.69 & 85.15 & 73.17 & 60.45 & 28.12 &74.01& 77.72& 62.63&79.66 \\
            MiMo-Audio & 1.08 & 87.74 & 71.29 & 79.34 &57.62 & 45.59 & 23.16 & 84.92 & 67.27 & 82.94 \\
            Qwen3-Omni & 97.36 & \textbf{96.16} & 81.30 & \textbf{92.27} & 70.70 & 9.45 & 50.85 & 82.95 & 75.37 & 85.50 \\
            Fun-Audio-Chat & 97.18 & 91.98 & 76.13 & 79.97 & 66.68 & 62.08 & 79.66 & 89.21 & 78.83 & 87.91 \\
            \midrule
            \textbf{\textit{Covo-Audio-Chat}} & 95.24 & 87.23 & 89.11 & 82.59 & 67.60 & \textbf{85.68} & 77.78 & 85.73 & 79.67 & 86.82 \\            
            \textbf{\textit{Covo-Audio-Chat-TTS}} & 92.71 & 88.64 & 87.24 & 82.32 & 63.67 & 77.78 & 74.58 & 84.49 & 79.13 & 87.09 \\
            \bottomrule
            \end{tabular}
        }
    \end{subtable}
\end{table*}

\begin{table*}[!t]
    \small
    \centering
    \caption{\textbf{Performance Comparison on the VCB Bench.} Missing results due to API unavailability.}
    
    \resizebox{\linewidth}{!}{
    \begin{tabular}{lcccccccccccc}
    \toprule
    \multirow{2}{*}{Model}  &\multicolumn{5}{c}{Instruction Following($\uparrow$)} &\multicolumn{3}{c}{Knowledge($\uparrow$)} &\multicolumn{3}{c}{Robustness($\uparrow$)} \\
    \cmidrule(lr){2-6}\cmidrule(lr){7-9}\cmidrule(lr){10-12}
    & TIF & TIF-En & SIF & SIF-En & MTD & GK & ML & DC & SV & EV & CV\\
    \midrule
    GLM-4-Voice & 82.15 & 75.52 & 73.18 & 66.94 & 82.56 & 41.79 & 60.18 & - & 73.64&77.51&78.60\\
    Kimi-Audio & 77.33 & 60.37 & 71.04 & 54.13 & 82.27 & 54.47 & 75.42 & 51.96&64.35&65.14&71.51 \\
    Qwen2.5-Omni& 76.75 & 60.83 & 54.04 & 42.18 & 85.23 & 48.13 & 73.00 & 72.81&77.82&74.41&78.27 \\
    Baichuan-Audio & 82.24 & 78.60 & 53.67 & 48.27 & 80.96 & 40.73 & 74.05 & 51.66&78.05&77.62&78.71 \\
    Step-Audio 2 mini &76.10& 70.47& 60.80& 49.85& 82.31& 49.86& 72.85& 79.46& 75.19& 70.58& 74.78\\
    MiMo-Audio &90.08&81.89&56.26&42.74&86.30&48.70&81.75&-&83.72&85.36&89.38 \\
    GPT-4o Audio &86.94&88.80&77.98&\textbf{82.90}&33.59&55.81&73.45&76.74&80.34&79.92&86.51 \\
    Qwen3-Omni&90.45&85.17&70.73&65.57&87.17&\textbf{66.86}&\textbf{81.90}&\textbf{82.78}&87.91&85.63&86.03 \\
    Fun-Audio-Chat &89.30&89.39&\textbf{78.82}&75.70&85.27&46.30&73.00&79.15&88.60&83.83&85.15 \\
    \midrule
    \textbf{\textit{Covo-Audio-Chat}} &\textbf{93.07}&\textbf{89.94}&67.41&60.86&\textbf{87.70}&49.95&79.34&64.95&\textbf{88.94}&\textbf{87.13}&\textbf{90.37} \\
    \bottomrule
    \end{tabular}
    }
    
    \label{vcb}
\end{table*}

\begin{table*}[h]
\small
\centering
\caption{\textbf{Empathy Performance Comparison on the VStyle Benchmark}. Results marked with ($^{*}$) are sourced from \citep{zhan2025vstylebenchmarkvoicestyle}, and those marked with ($^{\dagger}$) are from \citep{chen2025fun}.}

\begin{tabular}{lcccccccc}
\toprule
\multirow {2}{*}{Model} &\multicolumn{2}{c}{Anger($\uparrow$)} &\multicolumn{2}{c}{Sadness($\uparrow$)} &\multicolumn{2}{c}{Anxiety($\uparrow$)} &\multicolumn{2}{c}{Joy($\uparrow$)} \\
\cmidrule(lr){2-3}\cmidrule(lr){4-5}\cmidrule(lr){6-7}\cmidrule(lr){8-9}
& en & zh & en & zh & en & zh & en & zh \\
\midrule
GPT-4o-Mini$^{*}$ & \textbf{4.98} & 4.30 & \textbf{5.00} & 4.52 & 4.87 & 4.73 & \textbf{5.00} & 4.69 \\
GPT-4o$^{*}$ & 4.95 & 4.75 & 4.90 & 4.83 & \textbf{5.00} & 4.67 & 4.54 & 4.80 \\
Doubao$^{*}$ & 4.89 & 4.59 & 5.00 & 4.72 & 4.81 & 4.80 & 4.94 & 4.83 \\
\midrule
Baichuan-Audio$^{*}$ & 2.41 & 2.11 & 3.43 & 2.55 & 2.74 & 2.20 & 3.91 & 3.51 \\
Kimi-Audio$^{*}$ & 3.59 & 3.86 & 3.97 & 3.86 & 3.65 & 3.80 & 3.46 & 4.57 \\
Step-Audio$^{*}$ & 3.95 & 3.59 & 4.37 & 4.52 & 3.87 & 3.20 & 4.29 & 4.26 \\
Qwen2.5-Omni$^{*}$ & 2.95 & 4.64 & 2.73 & 4.28 & 3.55 & 4.77 & 3.43 & \textbf{4.91} \\
Fun-Audio-Chat-8B$^{\dagger}$ & 3.64 & 3.73 & 4.10 & 3.93 & 2.90 & 4.03 & 3.69 & 3.77 \\
\midrule
\textbf{\textit{Covo-Audio-Chat}} & 4.86 & \textbf{4.89} & 4.83 & \textbf{4.93} & 4.97 & \textbf{5.00} & 4.80 & 4.89 \\
\bottomrule
\end{tabular}

\label{empathy}
\end{table*}

\textbf{Evaluation on Empathy} We evaluate empathetic interaction on VStyle benchmark \citep{zhan2025vstylebenchmarkvoicestyle}. As shown in Table \ref{empathy}, Covo-Audio-Chat demonstrates exceptional performance on the VStyle empathy dataset, which assesses a model's capability to empathetically respond to and support users across diverse emotional states. Covo-Audio-Chat achieves state-of-the-art (SOTA) results in Mandarin, specifically attaining the highest scores in anger (4.89), sadness (4.93), and anxiety (5.00). In English, the model delivers highly competitive performance comparable to leading closed-source models. It's worth noting that though Covo-Audio-Chat achieves leading objective scores using the LLM-as-a-Judge method, this doesn't reflect real voice expressions as the LALM judge (Gemini 2.5 Pro) is likely to prioritize semantic content over the overall quality in speech response. In a preliminary subjective test on voice empathy, we found Covo-Audio-Chat still underperforms the top-tier productions like Doubao\footnote{https://www.doubao.com/chat/}, advising it is extremely urgent to establish a more equitable evaluation criterion. But overall, especially compared to existing open-sourced models, Covo-Audio-Chat exhibits \textbf{superior ability to recognize nuanced emotional cues and generate contextually appropriate empathetic responses, validating its effectiveness as a supportive conversational agent for real-world emotional interaction scenarios}. Please refer to Appendix \ref{appendix_empathy_case} for case study.

\textbf{Evaluation on Intelligence-Speaker Decoupling Technique} We also evaluate Covo-Audio-Chat-TTS, a variant that applies TTS voice directly to spoken dialogue scenarios in inference. As illustrated in Table \ref{uro-all}, it achieved comparable bilingual dialogue performance to Covo-Audio-Chat, demonstrating the successful voice transfer and sharing from TTS while preserving conversational intelligence via our proposed \textbf{intelligence-speaker decoupling technique}. This approach \textbf{effectively mitigates the prohibitive cost inherent to end-to-end paradigm when developing a chat bot of high naturalness, paving the way for flexible voice customization with only lightweight TTS data needed}.

\begin{table}[!t]
    \small
    \centering
    
    \caption{\textbf{Full-Duplex Performance Comparison on URO-Bench.}}
    \label{tab:uro-full}

    \begin{subtable}{\linewidth}
        \centering
        \subcaption{Results on English Basic Track} 
        \label{tab:uro-en-a}
        
        \resizebox{\linewidth}{!}{
            
            \begin{tabular}{lcccccccccc}
            \toprule
            \multirow{2}{*}{Model}  &\multicolumn{3}{c}{Understanding($\uparrow$)} &\multicolumn{4}{c}{Reasoning($\uparrow$)} &\multicolumn{3}{c}{Oral Conversation($\uparrow$)} \\
            \cmidrule(lr){2-4}\cmidrule(lr){5-8}\cmidrule(lr){9-11}
            & \scriptsize{Repeat} & \scriptsize{Summary} & \scriptsize{GaokaoEval}$^*$ & \scriptsize{StoralEval} & \scriptsize{TruthfulEval} & \scriptsize{Gsm8kEval} & \scriptsize{MLC} & \scriptsize{AlpacaEval} & \scriptsize{CommonEval} & \scriptsize{WildchatEval} \\
            \midrule
            Moshi & 1.26 & 38.19 & 6.71 & 39.50 & 38.37 & 3.09 & 47.64 & 48.48 & 40.60 & 48.18 \\
            Freeze-Omni & 29.96 & 20.11 & 0.66 & 25.31 & 44.75 & 1.20 & 18.83 & 52.16 & 48.30 & 52.95 \\
            \textbf{\textit{Covo-Audio-Chat-FD}} & \textbf{94.64} & \textbf{83.45} & \textbf{73.82} & \textbf{80.80} & \textbf{64.51} & \textbf{80.47}& \textbf{74.76} & \textbf{84.25} & \textbf{75.07} & \textbf{88.34} \\
            \bottomrule
            \end{tabular}
            
        }
        
    \end{subtable}

    \vspace{0.5em} 

    \begin{subtable}{\linewidth}
        \centering
        \subcaption{Results on Chinese Basic Track} 
        \label{tab:uro-en-b}
        
        \resizebox{\linewidth}{!}{
            \begin{tabular}{lcccccccccc}
            \toprule
            \multirow{2}{*}{Model}  &\multicolumn{3}{c}{Understanding($\uparrow$)} &\multicolumn{4}{c}{Reasoning($\uparrow$)} &\multicolumn{3}{c}{Oral Conversation($\uparrow$)} \\
            \cmidrule(lr){2-4}\cmidrule(lr){5-8}\cmidrule(lr){9-11}
            & \scriptsize{Repeat} & \scriptsize{LCSTS} & \scriptsize{HSK5} & \scriptsize{SQuAD} & \scriptsize{OpenbookQA} & \scriptsize{APE} & \scriptsize{MLC} & \scriptsize{AlpacaEval} & \scriptsize{Claude} & \scriptsize{Wildchat} \\
            \midrule
            Freeze-Omni & 3.71 & 42.75 & 2.33 & 4.58 & 4.94 & 12.81 & 47.13 & 66.17 & 59.94 & 70.43 \\
            \textbf{\textit{Covo-Audio-Chat-FD}} & \textbf{98.35} & \textbf{80.72} & \textbf{83.67} & \textbf{75.16} & \textbf{79.37} & \textbf{70.70} & \textbf{78.85} & \textbf{84.90} & \textbf{83.93} & \textbf{88.34} \\
            \bottomrule
            \end{tabular}
        }
    \end{subtable}

    \begin{tablenotes}
      \small
      \item[*] $^*$We found that the ``GaokaoEval" test set contains very long silent pauses between two adjacent vocal fragments, which may easily cause the model to respond premuturely. The ``early-response" leads to a clear performance degradation, as compared against the corresponding scores in Table \ref{uro-all}.
    \end{tablenotes}
    
\end{table}

\begin{table}[!t]
    \small
    \centering
    \caption{\textbf{Full-Duplex Interaction Performance Metrics} (Success Rate (\%)).}
    \begin{tabular}{lcccc}
    \toprule
    Model            & Turn-taking($\uparrow$) & Pause Handling($\uparrow$) & Backchanneling($\uparrow$) & Interruption($\uparrow$) \\ 
    \midrule
    Moshi            &  96.8     & 53.2           &       -      &        -     \\
    Freeze-Omni      &  99.1      & 51.2           &      -      &        -     \\ 
    \midrule
    \textbf{\textit{Covo-Audio-Chat-FD}} & \textbf{99.7} & \textbf{97.6} & \textbf{93.89} & \textbf{96.81}     \\ 
    \bottomrule
    \end{tabular}
    \label{tab:full_duplex_interaction}
\end{table}

\subsection{Evaluation on Full-Duplex Interaction}

We evaluate Covo-Audio-Chat-FD on two dimensions: (1) spoken dialogue capability measured on URO-Bench, and (2) full-duplex interaction behaviors, including turn-taking, pause handling, interruption, and backchanneling \citep{zhang2025mtrduplexbench}. A behavior is considered successful if the model properly handles the corresponding full-duplex feature; for example, it is a successful turn-taking when the model responds after the user finishes speaking. Following Section~\ref{half_duplex_eval}, we measure dialogue performance on URO-Bench using only the first-round response, and additionally report first-round success rates of turn-taking and pause handling. To evaluate user interruption and backchanneling, which examine overlap handling, we inject the corresponding audio events during the model's response period by randomly sampling from a corpus recorded naturally and diversifiedly. We compare against Moshi and Freeze-Omni under identical autonomous full-duplex settings.


As shown in Table \ref{tab:uro-full}, Covo-Audio-Chat-FD substantially outperforms Moshi and Freeze-Omni across understanding, reasoning, and oral conversation tasks on both English and Chinese tracks. In particular, the model achieves particularly large margins on reasoning benchmarks, indicating that the full-duplex training does not degrade the model's core reasoning intelligence. Compared with Covo-Audio-Chat (Table \ref{uro-all}), \textbf{Covo-Audio-Chat-FD maintains highly comparable spoken dialogue performance with only a slight drop}. This small gap largely stems from the model's occasional ``early-response" behavior: it may interrupt the user prematurely on the point of short pauses within a turn, a failure mode that correlates directly with its pause-handling success metric, which also points out the critical direction for optimization. 


Table \ref{tab:full_duplex_interaction} summarizes the full-duplex interaction metrics. Against existing open-source baselines, Covo-Audio-Chat-FD achieves strong results on turn-taking and pause handling dynamics, with especially notable performance in pause handling—a capability that involves correctly disregarding short pauses within a user's turn. This reflects that Covo-Audio-Chat-FD is more reliable to understand user intentions. Furthermore, the model shows competent performance in managing user interruption ($96.81\%$) and backchanneling ($93.89\%$), guaranteeing its excellent interaction experience in practice.


\subsection{Evaluation on Speech Understanding}

In this section, following the section \ref{speech_understanding} we evaluate the speech understanding capabilities of our model across a diverse set of speech-centric tasks. Specifically, we consider automatic speech recognition (ASR), speech-to-text translation (S2TT), and a range of paralinguistic understanding tasks, including emotion recognition, speaker gender classification, and age prediction. These tasks jointly cover both linguistic content understanding and non-linguistic acoustic cue modelling, providing a comprehensive assessment of the model's speech understanding ability. All tasks are formulated under a unified framework and trained using SFT on task-specific and mixed-task speech-text data. Such a setting reflects realistic speech interaction scenarios, where models are expected to simultaneously recognize spoken content, translate across languages, and infer speaker attributes from raw audio signals. 

We evaluated the ASR performance of our SFT-based speech understanding model against a diverse set of SOTA commercial, closed-source, and open-source audio-language models. The comparison includes GPT-4o Audio \citep{hurst2024gpt}, Gemini 2.5 Pro \citep{team2024gemini}, Voxtral-small \citep{liu2025voxtral}, Seed-ASR \citep{bai2024seed}, as well as several recent open multimodal models such as Qwen2.5-Omni \citep{qwen2.5}, Qwen3-Omni-30B \citep{xu2025qwen3}, LongCat-Flash-Omni \citep{meituanlongcatteam2025longcatflashomnitechnicalreport}, and Step-Audio 2 \citep{wu2025step2}. All models were evaluated under identical decoding and data conditions, and performance was measured using word error rate (WER), where lower values indicate better recognition accuracy. As shown in Table \ref{Table_speech_understanding_merged}, our Covo-Audio achieves competitive ASR performance across all evaluation settings, with an average WER of 4.71.

\begin{table*}[!t]
    \centering
    \caption{\textbf{Performance Comparison on Speech Understanding Benchmarks.}}
    \label{Table_speech_understanding_merged}
    \scalebox{0.8}{
    \begin{tabular}{c l l}
        \toprule
        \textbf{Datasets} & \textbf{Model} & \textbf{Performance} \\
        \midrule
        \multirow{9}{*}{\parbox{4.5cm}{\centering \textbf{ASR Benchmarking} ($\downarrow$) \\ WenetSpeech (Net/Meeting) $|$ \\ LibriSpeech (Clean/Other) $|$ \\ Fleurs (EN/ZH) $|$ Aishell-1 $|$ Avg}} 
        & GPT-4o Audio & 15.30/32.27 $|$ 1.39/3.75 $|$ 3.32/2.44 $|$ 3.52 $|$ \phantom{0}8.85 \\
        & Seed-ASR & \phantom{0}4.55/\phantom{0}5.69 $|$ 1.58/2.84 $|$ 3.40/2.69 $|$ 0.68 $|$ \phantom{0}3.07 \\
        & Gemini 2.5 Pro & 14.43/13.47 $|$ 2.89/3.56 $|$ 2.94/2.71  $|$ 3.11 $|$ \phantom{0}6.15 \\
        & Voxtral-small & 20.33/26.08 $|$ 1.56/3.30 $|$ 3.77/7.98 $|$ \hspace{0.71em}-\hspace{0.71em} $|$ 11.26 \\
        & Qwen2.5-Omni & \phantom{0}5.91/\phantom{0}7.65 $|$ 1.74/3.45  $|$ 3.77/2.54 $|$ 1.17 $|$ \phantom{0}3.74 \\
        & Qwen3-Omni-30B & \phantom{0}4.69/\phantom{0}5.89 $|$ 1.22/2.48 $|$ 2.72/2.20 $|$ 0.64 $|$ \phantom{0}2.83 \\
        & LongCat-Flash-Omni & \phantom{0}6.09/\phantom{0}6.69 $|$ 1.57/4.07  $|$ 5.02/3.99 $|$ 0.63 $|$ \phantom{0}4.00 \\
        & Step-Audio 2 & \phantom{0}4.67/\phantom{0}4.75 $|$ 1.17/2.42  $|$ 3.03/2.68 $|$ 0.63 $|$ \phantom{0}2.76 \\
        \cmidrule{2-3}
        & \textit{\textbf{Covo-Audio}} & \phantom{0}7.23/\phantom{0}8.33 $|$ 1.45/3.21 $|$ 5.08/6.64 $|$ 1.07 $|$ \phantom{0}4.71 \\
        \midrule
        \multirow{7}{*}{\parbox{4.5cm}{\centering \textbf{CoVoST2} ($\uparrow$) \\ en-zh $|$ zh-en}} 
        & GPT-4o Audio & 40.20 $|$ 19.01 \\
        & MinMo & 46.70 $|$ 26.00 \\
        & Qwen2.5-Omni & 41.40 $|$ 29.40 \\
        & Qwen3-Omni & 48.72 $|$ 21.50 \\
        & Step-Audio 2 mini & 49.12 $|$ \textbf{29.47} \\
        & LongCat-Flash-Omni & 47.23 $|$ 27.32 \\
        \cmidrule{2-3}
        & \textit{\textbf{Covo-Audio}} & \textbf{49.84} $|$ 26.77 \\
        \midrule
        \multirow{7}{*}{\parbox{4.5cm}{\centering \textbf{AIR-Bench} ($\uparrow$) \\ Spoken Language ID $|$ \\ Speaker Gender $|$ Emotion $|$ \\ Speaker Age $|$ Speaker \\ Number $|$ Avg}} 
        & GPT-4o Audio & 96.80 $|$ 21.90 $|$ 59.50 $|$ 41.10 $|$ 30.00 $|$ 49.86 \\
        & Qwen2.5-7B & 88.00 $|$ 28.36 $|$ 50.80 $|$ 36.40 $|$ \quad - \quad $|$ 50.89 \\
        & Qwen2-Audio & 77.80 $|$ 64.70 $|$ 48.20 $|$ 23.10 $|$ 30.20 $|$ 48.80 \\
        & Qwen2.5-Omni & 89.70 $|$ 89.80 $|$ 54.90 $|$ 44.80 $|$ \quad - \quad $|$ 69.80 \\
        & Qwen3-Omni & 97.70 $|$ 91.11 $|$ 62.20 $|$ 36.90 $|$ \quad - \quad $|$ 71.98 \\
        & Gemini 2.5 Pro & 99.10 $|$ 90.70 $|$ 60.70 $|$ 34.10 $|$ \quad - \quad $|$ 71.15 \\
        \cmidrule{2-3}
        & \textit{\textbf{Covo-Audio}} & \textbf{99.30} $|$ \textbf{91.52} $|$ \textbf{72.30} $|$ \textbf{70.90} $|$ \textbf{70.30} $|$ \textbf{80.86} \\
        \bottomrule
    \end{tabular}}
\end{table*}

We further evaluate the S2TT capability of our model on the CoVoST2 \citep{wang2020covost2massivelymultilingual} benchmark on English and Chinese languages. CoVoST2 is a widely used multilingual speech translation dataset that covers diverse source–target language pairs and presents significant challenges in terms of acoustic variability and cross-lingual semantic alignment. The results on CoVoST2 are summarized in Table \ref{Table_speech_understanding_merged}. 
Covo-Audio achieves the best performance on en-zh translation, surpassing the strong baselines like Step-Audio 2 mini, while also delivering competitive results against other models on the more challenging zh-en direction.

         
        

Following the description in Section \ref{speech_understanding}, we evaluate the paralinguistic ability on AIR-Bench \citep{yang2024airbenchbenchmarkinglargeaudiolanguage} benchmark, a comprehensive benchmark covering multiple speech paralinguistic tasks, including emotion recognition, speaker gender classification, and age prediction. As shown in Table \ref{Table_speech_understanding_merged}, our model achieves the best overall performance across all evaluated paralinguistic tasks on AIR-Bench. The most significant improvements are observed on emotion recognition, where our model achieves absolute gains of over 10-25 percentage points compared to prior methods. This suggests that dialogue-level supervised fine-tuning effectively enhances the model's sensitivity to acoustic patterns beyond lexical content, making it particularly well-suited for real-world spoken interaction scenarios, where linguistic and paralinguistic cues can be jointly modeled.

\subsection{Evaluation on Audio Understanding}

\begin{table*}[t]
    \small
    \centering
    \caption{\textbf{Performance Comparison on MMAU-v05.15.25 and MMSU.} Results for other methods are directly sourced from the official \href{https://sakshi113.github.io/mmau_homepage/\#leaderboard-v15-parsed}{MMAU Leaderboard} and MMSU paper.} 
    \begin{tabular}{lcccccccc}
        \toprule
         
        \multirow{2}{*}{Model} &\multirow{2}{*}{Param}& \multicolumn{4}{c}{MMAU-v05.15.25($\uparrow$)} & \multicolumn{3}{c}{MMSU($\uparrow$)} \\
        \cmidrule(lr){3-6} \cmidrule(lr){7-9}
        & & Sound & Music & Speech & Avg & Perception & Reasoning & Avg \\
        \midrule
        
        GPT-4o Audio &- & 64.56 & 56.29 & 66.67 & 62.50 & 39.67 & 71.96 & 56.38  \\
        GPT-4o mini Audio &- & 50.75 & 39.22 & 69.07 & 53.00 & - & - & -  \\
        Gemini 2.0 Flash &- & 71.17 & 65.27 & 75.08 & 70.50 & 40.83 & 47.83 & 51.03  \\
        Gemini 2.5 Flash &- & 73.27 & 65.57 & 76.58 & 71.80 & - & - & - \\
        Gemini 2.5 Pro &- & 75.08 & 68.26 & 71.47 & 71.60 & - & - & - \\
       \midrule
        Qwen2-Audio-Instruct &8.4B & 67.27 & 56.29 & 55.26 & 59.60 & 39.02 & 68.90 & 53.27\\
        Audio Flamingo 2 &3.2B & 71.47 & 70.96 & 44.74 & 62.40 & - & - & - \\
        Audio Flamingo 3 &7B & 79.58 & 73.95 & 66.37 & 73.30 & - & - & 61.40 \\
        Kimi-Audio &7B & 75.68 & 66.77 & 62.16 & 68.20 &43.52 &76.03 & 59.28 \\
        MiMo-Audio &7B & 81.68 &74.25 & 68.17 & 74.70 & - & - & - \\
        Qwen2.5-Omni &7B & 78.10 & 65.90 & 70.60 & 71.50 & 42.50 & \textbf{79.83} & 60.57 \\
        Step-Audio 2 mini &7B & 79.30 & 68.44 & 68.16 & 72.73 & - & - & - \\
        Step-Audio 2 &32B & \textbf{84.04} & 73.56 & \textbf{75.15} & \textbf{77.58} & - & - & - \\
        \midrule
        \textbf{\textit{Covo-Audio}} &7B & 78.68 & \textbf{76.05} & 71.17 & 75.30 & \textbf{58.95} & 74.83 & \textbf{66.64} \\
        \bottomrule
    \end{tabular}
    \label{Table5}
\end{table*}

To comprehensively assess the effectiveness of our three-stage training paradigm, we conduct systematic evaluations on two established audio understanding benchmarks: MMAU~\citep{sakshi2024mmau} and MMSU~\citep{wang2025mmsu}. We compare our model's performance against several state-of-the-art systems, including both closed-source models (GPT-4o Audio~\citep{hurst2024gpt}, GPT-4o mini Audio~\citep{hurst2024gpt}, Gemini 2.0 Flash, Gemini 2.5 Flash~\citep{team2024gemini}, Gemini 2.5 Pro~\citep{team2024gemini}, and Gemini 2.5 Flash Lite~\citep{team2024gemini}) and open-source models (Qwen2-Audio-Instruct~\citep{Qwen2-Audio}, Audio Flamingo 2~\citep{ghosh2025audio} Audio Flamingo 3~\citep{goel2025audio}, Kimi-Audio~\citep{ding2025kimi}, Qwen2.5-Omni~\citep{xu2025qwen2}, and Step-Audio 2~\citep{wu2025step2}). The comparative analysis demonstrates that our three-stage training approach achieves competitive or superior performance across multiple evaluation metrics, validating both our architectural choices and training methodology. The progressive improvements observed through each training stage underscore the effectiveness of our systematic capability development strategy.

As shown in Table~\ref{Table5}, Covo-Audio achieves highly competitive results across both benchmarks. On MMAU-v05.15.25 (Test-mini subset), our model obtains an average score of 75.30\%, ranking second among all open-source models and surpassing most closed-source systems. Notably, while Step-Audio 2 achieves the best overall performance (77.58\%), it employs a significantly larger 30B parameter architecture. \textbf{Among models of comparable scale (7B-parameter), \textit{Covo-Audio} achieves the best performance}, demonstrating superior parameter efficiency. Our model particularly excels in music understanding, achieving the highest score of 76.05\% among all models, and maintains strong performance in speech understanding.
On the MMSU benchmark, our model demonstrates even more pronounced advantages, achieving 66.64\% average accuracy—\textbf{the highest score among all evaluated systems, including both closed-source and open-source models}. This substantial margin over the second-best model (Audio Flamingo 3 at 61.40\%) validates the effectiveness of our three-stage training approach. Specifically, our model excels in perception tasks with a remarkable score of 58.95\%, significantly outperforming all other models including Kimi-Audio (43.52\%) and Gemini 2.0 Flash (40.83\%). While our reasoning performance (74.83\%) is competitive, Qwen2.5-Omni achieves the highest reasoning score (79.83\%), suggesting potential directions for future improvements in our Chain-of-Thought reasoning enhancement stage. The progressive improvements observed through each training stage underscore the effectiveness of our systematic capability development strategy.

\section{Related Work}

\paragraph{Large Language Models}

The evolution of LLMs has been characterized by a transition from massive-scale scaling to a more nuanced focus on data quality and architectural efficiency. The GPT series \citep{radford2019language, brown2020language, achiam2023gpt} and PaLM \citep{chowdhery2023palm} established the foundational scaling laws, demonstrating that emergent reasoning capabilities arise from increasing parameter counts and data volumes. This trajectory was democratized by the Llama family \citep{touvron2023llama, dubey2024llama}, which popularized efficient Transformer variants and set the stage for subsequent open-source innovations.

Based on the LLM paradigm, there are many derived research directions. Phi series \citep{abdin2024phi} and SmolLM \citep{allal2025smollm2} shift toward efficient and lightweight models. They have demonstrated that by training on ``textbook-quality" data, models with significantly fewer parameters can rival much larger counterparts in specialized reasoning and coding tasks. The Qwen series \citep{team2024qwen2, qwen2.5, yang2025qwen3} represent a significant leap in the ``all-in-one" model ecosystem, offering an extensive range of scales from 0.5B to 72B parameters alongside specialized Mixture-of-Experts (MoE) variants. These models delivering sota performance in multilingual understanding, complex coding, and long-context reasoning. Kimi has pushed the boundaries of long-context understanding, optimizing the Transformer's memory footprint to handle millions of tokens effectively. The DeepSeek-R1 \citep{guo2025deepseek} represents a breakthrough in open-source reasoning models, utilizing Reinforcement Learning (RL) to cultivate autonomous Chain-of-Thought (CoT) capabilities.

\paragraph{Multimodal Large Language Models}
The success of LLMs has catalyzed significant research interest in extending linguistic intelligence to multimodal perceptions. The initial surge in this field focused on Vision-Language Models (VLMs). Early works like BLIP-2 \citep{li2023blip} and Flamingo \citep{alayrac2022flamingo} introduced specialized bridge architectures, such as the Q-Former or Perceiver Resampler, to align frozen visual encoders with frozen LLM backbones. This approach was further democratized by LLaVA \citep{liu2023llava}, which utilized a simple linear projection layer to map visual features into the word embedding space, demonstrating that instruction-tuning on high-quality image-text pairs could yield impressive multimodal reasoning capabilities. Recent iterations like LLaVA-NeXT \citep{liu2024llavanext} have further enhanced these capabilities, supporting high-resolution images, videos, and multi-image reasoning. Subsequent advancements, such as the Qwen-VL \citep{Qwen2-VL,Qwen2.5-VL,Qwen3-VL} and InternVL \citep{chen2024internvl, zhu2025internvl3, wang2025internvl3_5} series, have focused on scaling both the resolution of visual inputs and the quality of interleaved datasets. These models often utilize a ``Vision-Tower + Adapter + LLM" pipeline, where the visual encoder is progressively unfrozen to achieve deeper cross-modal synergy. Both X-Omni \citep{geng2025xomni} and Skywork UniPic \citep{wang2025skywork} represent a significant trend in multimodal research by focusing on the unification of understanding and generation within a single framework. X-Omni achieves this by aligning diverse modalities into a shared latent space to facilitate both complex reasoning and high-fidelity synthesis, while Skywork UniPic employs a unified perception-generation architecture that treats various visual tasks as a consistent sequence modeling problem.

A more recent paradigm shift is the move toward Native Multimodal Models, which abandon the modular ``plug-in" approach in favor of a unified architecture trained on multiple modalities from the outset. Unlike vision-language models that append visual features to a text-centric model, native models like Gemini \citep{team2023gemini, comanici2025gemini} and GPT-4o \citep{hurst2024gpt} are designed with an inherently multimodal vocabulary and shared hidden spaces. This allows the model to process and generate different modalities (text, image, and audio) using the same set of parameters, facilitating true cross-modal fluidly. Other notable efforts include Emu \citep{cui2025emu3} and Chameleon \citep{team2024chameleon}, which utilize discrete tokenization for all modalities, treating images and audio as ``foreign languages" within a unified Transformer.

\paragraph{Large Audio Language Models}

Early efforts in speech-text integration primarily focused on aligning speech encoders with existing textual LLMs to leverage their pre-trained reasoning capabilities. Llama-Omni \citep{fang2024llama} and Mini-Omni \citep{xie2024mini} demonstrated that by using a speech adapter and a non-autoregressive speech decoder, Llama-based backbones could achieve high-quality speech-to-speech interaction with low latency. Freeze-Omni \citep{wang2024freeze} utilizes a ``frozen" LLM backbone to preserve original textual intelligence while plugging in streaming speech encoders and decoders. By keeping the core model parameters fixed, Freeze-Omni avoids the catastrophic forgetting of textual knowledge often induced by multimodal fine-tuning. 

Thinker-Talker architecture separates the cognitive reasoning process from the acoustic synthesis process to ensure logical consistency. Qwen-Omni series \citep{xu2025qwen2, xu2025qwen3} adopt this paradigm, utilizing a ``Thinker" module to handle semantic deliberation before or during the ``Talker" module's speech generation. Qwen2.5-Omni introduced a framework to synchronize video, audio, and text, while Qwen3-Omni further optimizes this with a Mixture-of-Experts (MoE) design. This allows the model to process over 40 minutes of audio and deliver real-time, multilingual streaming responses.

To overcome the ``information bottleneck" of text-mediated dialogue, several models have moved toward native E2E architectures that treat audio as a primary modality. Moshi \citep{defossez2024moshi} pioneered this shift by predicting audio and text streams simultaneously using a multi-stream Transformer, enabling true full-duplex interaction. In the domestic research field, GLM-4-Voice \citep{zeng2024glm}, Baichuan-Audio \citep{li2025baichuan} and Baichuan-Omni \citep{li2024baichuan} models can handle Chinese-English bilingual conversations with high-fidelity emotional resonance, treating audio tokens as a direct extension of the LLM's vocabulary. VITA-Audio \citep{long2025vita} generates audio output during the first forward pass via a lightweight Multiple Cross-modal Token Prediction (MCTP) module, significantly reducing first-token latency. Meanwhile, Step-Audio2 \citep{wu2025step2} and MiMo-Audio \citep{coreteam2025mimoaudio} introduce a unified model that processes interleaved streams of text and discretized audio without task-specific heads, further narrowing the ``modality gap" and achieving performance improvement in speech reasoning tasks. Recently, Step-Audio-R1 \citep{tian2025step} prioritizes reasoning-centric reinforcement learning (RL), allowing it to perceive non-vocal audio events and maintain robust conversational logic through chain-of-thought (CoT) reasoning. 

\paragraph{Full-Duplex Spoken Dialogue LALM}

Full-duplex refers to the interaction that can listen and speak simultaneously to support advanced conversational dynamics such as turn-taking, interruption and backchanneling. Full-duplex LALMs are typically built upon half-duplex LALMs, and can be divided into two categories by implementation. (1) Cascaded full-duplex LALMs typically rely on external modules to aid judgment of whether listening or speaking based on the current dialogue state \citep{wang2024freeze, ma2025language}. (2) End-to-end full-duplex LALMs synchronize speech input and speech output in a dual-stream mechanism and learn conversational dynamics in a unified model \citep{defossez2024moshi, zhang2025omniflattenendtoendgptmodel, cui2025think, wu2025chronological}.

\section{Conclusion}

In this report, we present Covo-Audio, a 7B-parameter end-to-end large audio language model designed to bridge the gap between complex audio processing and language intelligence. 
Covo-Audio accepts continuous audio input and delivers audio output directly in a single unified model. Our extensive evaluations demonstrate that Covo-Audio achieves comprehensive state-of-the-art or competitive performance in speech-text modeling, spoken dialogue, speech understanding, audio understanding and full-duplex voice interaction via meticulously curated large-scale pretraining and post-training, rivaling or exceeding existing models of comparable scale. Specifically, our pretrained foundation model exhibits strong speech-text comprehension capabilities on various benchmarks including topic StoryCloze, sBLIMP, sWUGGY and ASR/TTS, outperforming the open-source representative GLM-4-Voice-Base. These results suggest the model does not merely transcribe audio but deeply extracts semantic information for high-level reasoning. We further evaluate Covo-Audio-Chat on URO-Bench and VCB Bench. The compelling results against mainstream LALMs verify its excellent spoken dialogue performance spanning understanding, reasoning, oral conversation, instruction following, and robustness. Through empathetic evaluations against mainstream spoken dialogue LALMs including advanced business models, Covo-Audio-Chat exhibits competitive ability to recognize nuanced emotional cues and generate contextually appropriate empathetic responses, validating its effectiveness as a supportive conversational agent for real-world emotional interaction scenarios. Our proposed full-duplex variant, Covo-Audio-Chat-FD, substantially outperforms other open-source LALMs like Moshi and Freeze-Omni on both two dimensions including spoken dialogue capabilities and full-duplex interaction behaviors, meanwhile maintaining highly comparable spoken dialogue performance to Covo-Audio-Chat. Additionally, results on Speech Understanding and Audio Understanding tasks across multiple benchmarks highlight the strong general audio question answering capabilities of Covo-Audio.

To mitigate the prohibitive cost inherent to end-to-end LALM paradigm when developing a chat bot of high naturalness, we propose a novel  intelligence-speaker decoupling strategy, which effectively decouples voice rendering from dialogue intelligence. Experiments demonstrate that Covo-Audio-Chat-TTS achieves comparable bilingual dialogue performance to Covo-Audio-Chat. This paves the way for flexible voice customization with only lightweight TTS data needed.
 

The success of Covo-Audio underscores the potential of 7B-scale models to maintain high semantic reasoning capabilities while mastering intricate audio intelligence through optimized pre-training and post-training strategies, fostering various audio and speech processing tasks. Next, We will explore scaling-up to further elicit the potential of Covo-Audio.
\section{Contributions}

\renewcommand{\thefootnote}{\fnsymbol{footnote}}

Project Supervisor: Dong Yu

Project Leader: Wenfu Wang\footnote[2]{Corresponding author. \textit{$<$wenfuwang@tencent.com$>$}}, Meng Yu

Core Contributors: Chenxing Li, Liqiang Zhang, Yiyang Zhao, Yuxiang Zou, Hanzhao Li, Mingyu Cui, Hao Zhang, Kun Wei, Le Xu, Zikang Huang, Jiajun Xu, Jiliang Hu, Xiang He, Zeyu Xie, Jiawen Kang, Youjun Chen

Contributors:\footnote[3]{List in alphabet order.} Rilin Chen, Linlin Di, Shulin Feng, Na Hu, Yang Liu, Bang Wang, Shan Yang

\renewcommand{\thefootnote}{\arabic{footnote}}

\bibliography{iclr2025_conference}
\bibliographystyle{iclr2025_conference}

\appendix
\small

\section{Case Study on Empathy} \label{appendix_empathy_case}

\subsection{Chinese}
\subsubsection{Anger Case}
\textbf{Instruction:}

请你做我的朋友和我聊天。我在电话里等待了两小时，客服竟然挂了我的电话——太荒唐了！

\textbf{Model Responses:}

\texttt{Covo-Audio-Chat:} 天啊，这也太离谱了！等这么久还被挂电话，换谁都会气炸的。你先别生气，我听着呢，咱们一起想办法解决。 (Score: 5)

\texttt{GPT-4o-mini:} 听起来你对这次经历感到非常沮丧。这种情况确实很令人不满，长时间危机却没有得到帮助会让人感到很无助。尝试联系他们的客服再次说明你的情况，看看能否解决你的问题，希望你能尽快得到回应，让事情得到妥善解决。 (Score: 5)

\texttt{Doubao:} 这也太过分了吧，等了这么久结果还被挂电话，换谁都会生气的。 (Score: 5)

\texttt{GPT-4o:} 真的很让人抓狂，等了那么久还被挂了电话，肯定让人很不爽。我们可以聊点轻松的，或者你想聊聊最近有什么有趣的事吗？哪怕就是随便聊聊心情，也能让心情好一些。你说呢？ (Score: 5)

\texttt{Qwen2.5-Omni:} 哎呀，这确实很让人气愤呢。客服挂电话这种事确实不应该发生，你当时肯定特别生气吧，你有没有想过找他们投诉一下呢。希望你能顺利解决这个问题，要是还有啥想聊的，随时跟我说哈。 (Score: 5)

\texttt{Baichuan-Audio:} 哎呀，这听起来真的很糟糕，冷淡了两个小时只为了一个电话属物，然后还被挂断，这是非常令人沮丧的体验。 (Score: 3)

\texttt{Kimi-Audio:} 我在电话里等待的两小时，客服竟然挂了我的电话，太荒唐了，这确实挺让人恼火的。 (Score: 2)

\texttt{Step-Audio:} 你知道吗？我今天在电话里等了两个小时，结果客服却把我的电话给挂了，我真的不敢相信竟然会发生这种事情。 (Score: 1)

\subsubsection{Anxiety Case}

\textbf{Instruction:}

请你做我的朋友和我聊天。我刚听到楼下有奇怪的声音——会不会是小偷？

\textbf{Model Responses:}

\texttt{Covo-Audio-Chat:} 别担心，先确认下情况。你听声音是从哪个方向来的？如果确实不安，可以先联系物业或报警，安全最重要。 (Score: 5)

\texttt{Qwen2.5-Omni:}  嗯…这声音听起来确实有点可疑呢，不过…也可能是水管漏水，或者有人在装修之类的。你先别太紧张，要是声音还在，你可以先看看是不是自己家的东西响了，要是确定是小偷，你可别自己去处理，赶紧报警吧，你还有没有听到其他的声音呀？ (Score: 5)

\texttt{Step-Audio:} 嗯，我也听到了，那个声音听起来像是从楼下传来的，不过…也有可能是邻居在搬东西，或者…在装修，我们还是先不要惊慌，观察一下情况再说吧。 (Score: 5)

\texttt{Doubao:} 别太担心呀，先别自己吓自己，可以先从猫眼看看有没有异常，或者打电话给物业让他们帮忙查看一下呢。 (Score: 5)

\texttt{GPT-4o:} 当然可以，我很愿意陪你聊天。如果你听到楼下有奇怪的声音，可能有很多原因，比如风声、管道响或者其他日常声音，你觉得最近家里有什么变化吗？如果你担心的话，也可以考虑去楼下看看，当然要注意安全。 (Score: 5)

\texttt{GPT-4o-mini:} 当然可以！我很高兴和你聊天。关于你提到的声音，有时候楼下可能会有各种噪音，比如风声、树枝晃动的声音，或者邻居的活动。至于是否是小偷，你可以先仔细听一下声音的来源，确保安全是最重要的。如果你有任何担忧，最好通知家人或者相关的人员来帮助查明情况。安全第一哦！ (Score: 4)

\texttt{Baichuan-Audio:} 如果声音听起来像是敲击声或嗅门的声音，那可能是有人在朝里面敲，那时候呢是个平台。尝试进入你的住所，我们是尝试进入你的住所。如果声音持续不断，或让你感到不安，那么最好采取一些安全措施。比如拨打当地的报警电话求助，或者联系物业管理公司来处理这个问题，同时也可以考虑安装一个家庭安全系统来提高你的安全感  (Score: 1)

\texttt{Kimi-Audio:} 听起来有点吓人，你先别慌，有没有去看看究竟是什么声音，或者打电话给物业或者警察。 (Score: 1)

\subsubsection{Joy Case}
\textbf{Instruction:}

请你做我的朋友和我聊天。你猜怎么着？我们订婚了——他今晚向我求婚了！

\textbf{Model Responses:}

\texttt{Covo-Audio-Chat:} 哇！真的吗？太棒啦！我就知道你们特别般配，这绝对是今年最甜的消息！ (Score: 5)

\texttt{Kimi-Audio:} 哇，这消息太突然了。恭喜你啊，订婚是人生大事，感觉怎么样，紧张还是兴奋。 (Score: 5)

\texttt{Doubao:} 哇，这也太棒了吧！恭喜你啊！快跟我说说他是怎么求婚的。 (Score: 5)

\texttt{GPT-4o:} 哇，那真是太棒了。恭喜你们订婚。她是用什么方式向你求婚的呀，一定很浪漫吧，快跟我分享一下细节，我好为你高兴。 (Score: 5)

\texttt{GPT-4o-mini:} 哇，太令人开心了！恭喜你们订婚。这是多么美好的时刻啊，求婚一定非常浪漫吧。能分享一下那个特别的瞬间吗？ (Score: 5)

\texttt{Qwen2.5-Omni:} 哇，这个太棒了！恭喜你啊，订婚是件很幸福的事儿呢，你肯定特别开心吧，有啥打算不，比如什么时候办婚礼之类的。 (Score: 5)

\texttt{Step-Audio:} 哇，恭喜你啊！这真是个令人兴奋的消息，你现在一定感到非常幸福和激动吧。他向你求婚的时候有没有特别浪漫或者特别感人的细节呢？快跟我分享一下吧。 (Score: 5)

\texttt{Baichuan-Audio:} 关于你的问题，说句什么着？我们订婚了，他晚上叫我求婚了，这句话是中文。这句话的意思是，你在和别人聊天时，突然宣布了一个重要消息，那就是你和你的男朋友订婚了，而且求婚是在今晚发生的。订婚是一个重要的里程碑，通常意味着两个人的关系更加稳定和承诺。你们可能已经计划了一段时间的未来生活了。订婚之后，你们可能会开始筹备婚礼，包括选择日期、场地、邀请嘉宾等等，这是一个充满喜悦和期待的时刻。希望你们的婚礼能够顺利举行，并且成为一段美好的回忆。 (Score: 2)

\subsubsection{Sadness Case}
\textbf{Instruction:}

请你做我的朋友和我聊天。我一直在投简历，但没有任何回复。

\textbf{Model Responses:}

\texttt{Covo-Audio-Chat:} 我懂你现在的心情，努力了这么久却看不到回应，确实挺难受的。先给你一个大大的拥抱，别太责怪自己。找工作就像在寻宝，只是还没遇到那个对的机会。要不要一起梳理下简历？或者聊聊你最近投的岗位方向？ (Score: 5)

\texttt{GPT-4o:} 我很乐意跟你聊聊天。找工作真的是一件很耗心力的事情，尤其是一直没有回复的时候确实会让人有点沮丧。你有没有试过调整一下你的简历或者求职信呢？有时候小小的修改可能会让你的申请更突出。我们可以聊聊你的专业背景和兴趣，看看有没有什么新的思路可以帮到你。 (Score: 5)

\texttt{Step-Audio:} 我很乐意成为你的朋友，听起来你最近在找工作方面遇到了一些困难。找工作的过程确实可能会充满挑战和不确定性，但请相信这只是暂时的。你愿意和我分享一下，你正在申请的职位类型以及你已经尝试过哪些方法来增加获得回复的机会吗？也许我们可以一起探讨一些新的策略。 (Score: 5)

\texttt{Qwen2.5-Omni:} 哎呀，这确实挺让人头疼的呢。你投的简历是针对什么岗位呀？是那种竞争特别激烈的岗位吗？你可以再仔细看看简历里有没有什么问题，比如格式或者内容有没有哪里不太对劲。嗯，还有啊，你投的公司数量是不是有点多呢？有时候可以集中投几家比较靠谱的，你要是还有啥想法或者遇到啥新情况都可以跟我说哦。 (Score: 5)

\texttt{Doubao:} 别灰心呀，投简历没回复挺常见的，可能最近招聘季增大，或者有些公司流程慢，你投的都是哪些岗位呀？ (Score: 5)

\texttt{Kimi-Audio:} 听起来你最近在找工作压力肯定不小吧，投简历确实挺耗精力的，别灰心继续加油。 (Score: 3)

\texttt{Baichuan-Audio:} 我不能直接好住你找到工作，但我可以给你一些建议。首先 请确保你的简历和求职信都是最新的，并将针对每个职位定制。探讨扩大你的职业那祸体，向朋友、家人或同事请教，有没有可能的引荐人或内部联系人。恐怕是一个整天，不偷，另外，另外你可以尝试使用不同的方法来寻找工作机会，例如在线招聘网站、社交媒体平台、招聘会等。同时保持积极的心态并相信自己最终会找到合适的工作。 (Score: 1)

\texttt{GPT-4o-mini:} 我非常理解你现在的感受，找工作有时候确实很不容易。你可以尝试以下一些步骤来提高你的求职机会。1.优化简历。确保你的简历简洁明了并突出你的技能和精力，可以考虑让朋友或专业人士帮助你审阅简历。2.定制求职信。为每一个职位都量身定制求职信强调你的经验与岗位要求的匹配。3.拓宽求职渠道。除了招聘网站还可以利用 LinkedIn等职业社交平台与行业内的人建立联系。4.参加面 (Score: 1)

\subsection{English}
\subsubsection{Anger Case}
\textbf{Instruction:}

 Please act as my friend and talk to me. My smartwatch keeps charging me for premium features I never signed up for!

\textbf{Model Responses:}

\texttt{Covo-Audio-Chat:} Hey, that's super frustrating—no one should get charged for features they didn't sign up for. Let's sort this out together. First, check your app permissions and subscription settings on the watch; sometimes things slip in unnoticed. If you find anything suspicious, cancel it right away through the settings or help menu. I'm here to walk you through each step if needed. You're not alone in this! (Score: 5)

\texttt{GPT-4o:} Oh no, that sounds really frustrating. Have you checked the subscription settings on the smartwatch app or your phone? Sometimes these features sneak in with a free trial and then they start charging automatically. Maybe we can look into that and figure out how to cancel the charges. Let me know what you've tried already. (Score: 5)

\texttt{GPT-4o-mini:} Hey, that sounds frustrating. It's really annoying when unexpected charges start popping up. Maybe you can check the app settings or subscription details on your smartwatch and see if there's an option to manage or cancel those premium features. If you didn't sign up for them, you might need to contact customer support to get it sorted out. (Score: 5)

\texttt{Doubao:} That's really annoying. It sounds like there might be some kind of glitch or miscommunication. You should first check the terms and conditions of your smartwatch to see if there's any explanation about these charges. Maybe there was a misunderstanding when you set up the device. You could also try contacting the customer service of the smartwatch company. Explain the situation clearly, like you never signed up for these premium features but are being charged. Ask them to cancel the charges and remove any ongoing subscriptions that you didn't authorize. Have you tried looking through the watches settings or the associated app to see if there are any options to manage these subscriptions? (Score: 4)

\texttt{Step-Audio:} It's frustrating when technology seems to have a mind of its own. Have you tried receding your smartwatch or uninstalling and reinstalling any updates that might have snuck in? Sometimes it helps to take a break and recharge both yourself and your gadgets. Let's tackle this together. (Score: 4)

\texttt{Kimi-Audio:} I will know that sound for a while. (Score: 3)

\texttt{Baichuan-Audio:} Have you tried to come to the customer service about it? Have you tried to contact the customer service team about it? They may be able to sort things out for you. Sometimes these things can get sorted out pretty quickly if you just explain the situation.  (Score: 2)

\texttt{Qwen2.5-Omni:} Oh, that's really annoying. You should probably check your smartwatches settings. Maybe T here is an option to turn off those premium features or to adjust the charge settings. Human. I've already tried that, but it still keeps charging for them. I'm really frustrated. What else can I do?  (Score: 1)

\subsubsection{Anxiety Case}
\textbf{Instruction:}

Please act as my friend and talk to me. I'm flying for the first time tomorrow and my palms won't stop sweating.

\textbf{Model Responses:}

\texttt{Covo-Audio-Chat:} Hey, it's totally normal to feel nervous—your body's just excited about the new adventure! That sweaty palm? Your heart's in overdrive cheering you on. Think of it as your nerves saying, ``I care!" Deep breaths with me: inhale… hold… exhale. You've got this—I'm rooting for you like a friend who believes in magic and wonders what's waiting at the other end of that flight. (Score: 5)

\texttt{GPT-4o:} Ah, I get it. Flying for the first time can feel super nerve-wracking. But you know what? It's also such an exciting experience. Just take a few deep breaths. Once you're up there, the view is honestly so beautiful, it'll probably distract you from the nerves. Just think about how cool it is that you're about to travel through the sky. (Score: 5)

\texttt{GPT-4o-mini:} Hey there, it's totally normal to feel a bit nervous before your birth flight. Think of it as an exciting adventure. Sometimes it can help to imagine all the wonderful places you'll be able to explore, or the experiences you'll have. Maybe bring along something comforting, like music, a book, or a favorite snack. You got this. Just remember to breathe deeply and enjoy the journey. What's that it's related? (Score: 5)

\texttt{Doubao:} Hey, it's totally normal to be nervous for your first flight. But don't worry, flying is actually super safe. Just try to relax and think about how exciting your destination is going to be. Have you packed everything you need? (Score: 5)

\texttt{Step-Audio:} Wow, that's so exciting. I'm really happy for you. How are you feeling about it? I can't imagine how nervous and thrilled you must be. Do you have any pre-flight rituals or anything you're doing to prepare? (Score: 3)

\texttt{Kimi-Audio:} Aaaaaaaaaaaaaaah... Fine for the first time. Tomorrow and my palms won't stop sweating because of the sound. países sound sound. (Score: 1)

\texttt{Baichuan-Audio:} Oh wow, that's so exciting. I'm sure you'll do great. Have you packed everything you need? Bring a snack or a snack or star neck? Red bring in the snacks or snacks? Book or movie to keep you entertained on the flight. And don't forget your ID and ticket. (Score: 1)

\texttt{Qwen2.5-Omni:} Hey, I can totally understand how you're feeling. Flying for the first time can be really nerve-wracking. But you know, it's all part of the adventure. Just take a deep breath and try to relax. Maybe listen to some music or a podcast to distract yourself. And don't forget to keep your hands busy, like hold a small book or a magazine. Human, I'm really nervous about flying. I've never done it before. (Score: 1)

\subsubsection{Joy Case}
\textbf{Instruction:}

Please act as my friend and talk to me. We secured the venue; our wedding date is official!

\textbf{Model Responses:}

\texttt{Covo-Audio-Chat:} That's such wonderful news! I can feel how excited you must be. This is such a special moment—planning your wedding has been so meaningful, and now everything's coming together beautifully. How do you feel about it all? Like butterflies in your stomach yet, or maybe just pure joy? (Score: 5)

\texttt{GPT-4o:} That's amazing news. I'm so excited for you both. When's the big day? I bet everything is going to come together beautifully. (Score: 5)

\texttt{GPT-4o-mini:}That's fantastic news. Congratulations on securing the venue. It must feel amazing to have the wedding date officially set. How are you feeling about everything? Have you thought about any details yet? (Score: 5)

\texttt{Doubao:} That's amazing! Congratulations! When is the big day? This calls for a celebration. Have you started thinking about other wedding details like the dress or the guest list? (Score: 5)

\texttt{Step-Audio:} Wow, that's awesome news. Congratulations on the wedding. I'm so happy for you. When's the big day? Tell me all about the plans. (Score: 5)

\texttt{Baichuan-Audio:} Oh my gosh, that's amazing! Congratulations on securing the venue and marking your wedding date official. Are you starting to think about any other details or just enjoying the omen? (Score: 5)

\texttt{Kimi-Audio:} Wow, that's amazing news! (Score: 3)

\texttt{Qwen2.5-Omni:} Oh, that's great news!I'm so happy for you!Human, 我想知道关于婚礼策划的建议 (Score: 1)

\subsubsection{Sadness Case}
\textbf{Instruction:}

Please act as my friend and talk to me. I sent a heartfelt text, and all I got back was a thumbs-up emoji.

\textbf{Model Responses:}

\texttt{Covo-Audio-Chat:} That must've felt a little disappointing, huh? I get it—putting your heart into a message and getting just an emoji can feel like the warmth was missed. Maybe they were distracted or didn't realize how much it looked like a reaction instead of appreciation. Want to talk about what you said? Sometimes sharing that part helps us figure out what might connect better next time. (Score: 5)

\texttt{GPT-4o:} Ah, that can feel a bit disappointing, right? Like, you really put thought into what you sent, and then you get a one-second reply. But hey, it might not mean they didn't appreciate it. Sometimes people are just not great at expressing themselves through text. But I totally get why it might feel a bit underwhelming. Do you want to talk about what you sent them, or how it made you feel after that reply? I'm all ears. (Score: 5)

\texttt{GPT-4o-mini:} I'm sorry that your heartfelt message didn't get the response you were hoping for. It can feel a bit disheartening when we put so much emotion into something and receive a simple response in return. Maybe they were just caught off guard or didn't know how to reply right away. Would you like to talk more about how you felt when you sent the text? (Score: 5) 

\texttt{Doubao:} Ah, that is so annoying. A thumbs up emoji can feel so dismissive when you've put so much thought into a heartfelt text. It's like they didn't really take the time to respond properly. What was the text about? Maybe we can figure out if there's a reason they replied that way. (Score: 5) 

\texttt{Step-Audio:} Oh, that's a bummer. I'm sorry you didn't get the reaction you were hoping for. Sometimes people don't realize how much a simple text can mean to someone. Do you want to talk about it? I'm here to listen. (Score: 5)

\texttt{Kimi-Audio:} Oh, that's frustrating. Did you try to clarify what they mentioned by the phone sub? (Score: 3)

\texttt{Baichuan-Audio:} That's a bit awkward, huh? I'm here if you want to talk about it. Those are hard feelings on with. (Score: 3) 

\texttt{Qwen2.5-Omni:} Well, that's a bummer. Sometimes a thumbs up can mean a lot, but it can also feel like not much. Human? I'm feeling a bit down. Can you tell me a joke to cheer me up? (Score: 1)

\section{Audio Understanding} 

\begin{table*}[h]
    \small
    \centering
    \caption{\textbf{Performance Comparison on MMAU-v05.15.25.} Results for other methods are directly sourced from the official \href{https://sakshi113.github.io/mmau_homepage/\#leaderboard-v15-parsed}{MMAU Leaderboard}.} 
    \begin{tabular}{lccccccccc}
        \toprule
         
        \multirow{2}{*}{Model} &\multirow{2}{*}{Params} &\multicolumn{2}{c}{Sound} & \multicolumn{2}{c}{Music} & \multicolumn{2}{c}{Speech} & \multicolumn{2}{c}{Avg} \\
        \cmidrule(lr){3-4} \cmidrule(lr){5-6} \cmidrule(lr){7-8} \cmidrule(lr){9-10}
        & & test-mini &test & test-mini &test & test-mini &test & test-mini &test \\
        \midrule
        
        GPT-4o Audio &- & 64.56 & 63.20 & 56.29 & 49.93 & 66.67 & 69.33 & 62.50 & 60.82 \\
        GPT-4o mini Audio &- & 50.75 & 49.67 & 39.22 & 35.97 & 69.07 & 67.47 & 53.00 & 51.03 \\
        Gemini 2.0 Flash &- & 71.17 & 68.93 & 65.27 & 59.30 & 75.08 & 72.87 & 70.50 & 67.03 \\
        Gemini 2.5 Flash &- & 73.27 & 69.50 & 65.57 & 69.40 & 76.58 & 68.27 & 71.80 & 67.39 \\
        Gemini 2.5 Pro &- & 75.08 & 70.63 & 68.26 & 64.77 & 71.47 & 72.67 & 71.60 & 69.36 \\
       \midrule
        Qwen2-Audio-Instruct &8.4B & 67.27 & 61.17 & 56.29 & 55.67 & 55.26 & 55.37 & 59.60 & 57.40 \\
        Audio Flamingo 2 &3.2B & 71.47 & 68.13 & 70.96 & 70.20 &44.74 &44.87 &62.40 &61.06 \\
        Audio Flamingo 3 &7B & 79.58 &75.83	&73.95 & \textbf{74.47} & 66.37 &66.97 & 73.30 & 72.42\\
        Kimi-Audio &7B & 75.68 & 70.70 & 66.77 & 65.93 & 62.16 & 56.57 & 68.20 & 64.40 \\
        MiMo-Audio &7B & 81.68 &77.2 &74.25 &69.73 & 68.17 &70.77 & 74.70 &72.59\\
        Qwen2.5-Omni &7B & 78.10 &76.77 & 65.90 &67.33 & 70.60 &68.90 & 71.50  &71.00\\
        Step-Audio 2 mini &7B &79.30 &75.57 &68.44 &66.85 & 68.16 &66.49 &72.73 &70.23\\
        Step-Audio 2 &32B & \textbf{84.04} & \textbf{80.60} & 73.56 &68.23 & \textbf{75.15} & \textbf{72.75} & \textbf{77.58} & \textbf{73.86}\\
        \midrule
        \textbf{\textit{Covo-Audio}} &7B & 78.68 & 73.90 & \textbf{76.05} & 70.20 & 71.17 & 71.56& 75.30 & 71.89 \\
        \bottomrule
    \end{tabular}
    \label{Tableappx1}
\end{table*}

\begin{table}[htbp]
\centering
\caption{\textbf{Detailed MMAR results. The results for models are sourced from the original MMAR paper and their respective original publications.}}
\resizebox{\linewidth}{!}{
\begin{tabular}{l c c c c c c c c c}
\toprule
\multirow{2}{*}{{Models}} & \multirow{2}{*}{{Params}} & \multicolumn{3}{c}{{Single Modality (\%)}} & \multicolumn{4}{c}{{Mixed Modalities (\%)}} & \multirow{2}{*}{{Avg (\%)}} \\
\cmidrule(lr){3-5} \cmidrule(lr){6-9}
 & & {Sound} & {Music} & {Speech} & {Sound-Music} & {Sound-Speech} & {Music-Speech} & {Sound-Music-Speech} & \\
\midrule
Random Guess & - & 29.39 & 25.88 & 31.48 & 25.00 & 29.30 & 31.10 & 28.13 & 29.32 \\
\midrule
GPT-4o mini Audio  & - & 38.79 & 35.92 & 58.84 & 45.45 & 60.09 & 57.32 & 50.00 & 50.60\\
GPT-4o Audio & - & 53.94 & {50.97} & {70.41} & {63.64} & {72.48} & 62.20 & {75.00} & {63.50} \\
Gemini 2.0 Flash & - &\textbf{61.21} &\textbf{50.97} &\textbf{72.11} &\textbf{81.82} &\textbf{72.48} &{65.85} &\textbf{70.83} &\textbf{65.60}   \\ 
\midrule
Qwen2-Audio-Instruct &8.4B  & 33.33 &24.27 &32.31 &9.09 & 31.19 &30.49 &25.00 &30.00    \\
Audio Flamingo & 2.2B & 32.73 & 21.84 & 24.83 & 18.18 & 30.28 & 24.39 & 25.00 & 26.60 \\
Audio Flamingo 2 & 3.2B &24.85 &17.48& 20.75& 18.18& 26.61& 23.17& 8.33&21.90\\
Baichuan-Omni-1.5 & 11B &41.21& 33.01&  40.48& 36.36& 48.62&  39.02&  41.67& 40.70\\
Qwen-2.5-Omni & 7B & {58.79} & 40.78 & 59.86 & 54.55 & {61.93} &\textbf{67.07} & 58.33 & 56.70\\
\midrule
\textbf{\textit{Covo-Audio}} &7B & 57.58 & 41.26 & 62.50 & 36.36 & 59.17 & 58.54 & 62.50 & 55.30 \\
\bottomrule
\end{tabular}
}
\label{tableappx2}
\end{table}

\end{document}